\documentclass{aa} 
\usepackage{graphicx} 

\usepackage{natbib} 
 
\usepackage{xspace} 

\newcommand{\paw}{$pa_{\omega}$\xspace} 
\newcommand{\thw}{$\theta_{\omega}$\xspace} 
\newcommand{\fco}{$f_{\rm co}$\xspace} 
\newcommand{\amp}{$\mathcal{A}$\xspace} 
\newcommand{\lat}{$\ell$\xspace} 
\newcommand{\degre}{$^{\circ}$\xspace} 
\newcommand{\kms}{km\,s$^{-1}$\xspace} 
 
\begin{document} 
 
\title{Interferometric imaging of carbon monoxide in comet C/1995 O1 (Hale-Bopp): evidence for a strong rotating jet} 
\author{D. Bockel\'ee-Morvan 
\inst{1} \and F. Henry\inst{1} \and N. Biver\inst{1} \and J. 
Boissier\inst{2} \and P. Colom\inst{1} \and J. Crovisier\inst{1} 
\and D. Despois\inst{3} \and R. Moreno\inst{1} \and J. 
Wink\inst{2}\thanks{deceased}} \institute{Observatoire de Paris, 
F-92195 Meudon, France  \and IRAM, 300 rue de la Piscine, Domaine 
universitaire, F-38406, Saint Martin d'H\`eres, France \and  
Observatoire de Bordeaux, BP 89, F-33270 Floirac, France} 
\offprints{D. Bockel\'ee-Morvan, 
\email{dominique.bockelee@obspm.fr}} 
\date{\today } 
\titlerunning{CO in Hale-Bopp (\today )} 
\authorrunning{Bockel\'ee-Morvan et al.} 
 
\abstract 
{Observations of the CO $J$(1--0) 115 GHz and $J$(2--1) 230 GHz 
lines in comet C/1995 O1 (Hale-Bopp) were performed with the IRAM 
Plateau de Bure interferometer on 11 March, 1997. The observations 
were conducted in both single-dish (ON--OFF) and interferometric 
modes with 0.13 km s$^{-1}$ spectral resolution. Images of CO 
emission with 1.7 to 3\arcsec~angular resolution were obtained.} 
{The ON--OFF and interferometric spectra show a velocity shift 
with sinusoidal time variations related to the Hale-Bopp nucleus 
rotation of 11.35 h. The peak position of the CO images moves 
perpendicularly to the spin axis direction in the plane of the 
sky. This suggests the presence of a CO jet, which is active night 
and day at about the same extent, and is spiralling with nucleus 
rotation. The high quality of the data allows us to constrain the 
characteristics of this CO jet.  } 
{We have developed a 
3-D model to interpret the temporal evolution of CO spectra and maps. The CO coma is 
represented as the 
combination of an isotropic distribution and a spiralling gas jet, both of 
nucleus origin.} 
{Spectra and visibilities (the direct output of interferometric 
data) analysis shows that the CO jet comprises $\sim$ 40\% the 
total CO production and is located at a latitude $\sim$ 20\degre 
North on the nucleus surface. Our inability to reproduce all 
observational characteristics shows that the real structure of the 
CO coma is more complex than assumed, especially in the first 
thousand kilometres from the nucleus. The presence of another 
moving CO structure, faint but compact and possibly created by an 
outburst, is identified.} 
{} 
 

\maketitle 
 
\section{Introduction\label{intro}} 
 
Millimetre spectroscopy provided many insights into 
the composition and physical properties of cometary atmospheres. 
Many cometary parent molecules issued from the nucleus were identified 
with this technique. The high spectral resolution capabilities, and 
the possibility to observe several rotational lines belonging to 
the same molecule, allowed us to retrieve unique information concerning the 
velocity and temperature of the expanding coma, and the anisotropy of gas 
production at the nucleus surface. Because of the low, diffraction-limited 
spatial resolution provided by radio dishes at millimetric wavelengths, 
at best typically 10\arcsec, studies concerning the spatial distribution of 
parent molecules in the coma were sparse.

The exceptional brightness of comet C/1995 O1 (Hale-Bopp) near its perihelion 
on 1 April, 1997, motivated a wealth of innovative cometary observations. 
Among them, interferometric imaging of rotational transitions of parent 
molecules was successfully attempted. The Berkeley-Illinois-Maryland 
Association (BIMA) array mapped comet 
Hale-Bopp in HCN $J$(1--0) and CS $J$(2--1) with 9\arcsec~angular 
resolution. Spatial asymmetries suggesting the presence of gas 
jets were detected \citep{vea00,wri98,woo02}. 
Constraints on the photodissociative 
scalelengths of HCN and CS were obtained from the radial extent of 
their radio emissions \citep{sny01}. Using the Owens Valley 
Radio Observatory (OVRO) millimetre array, \citet{bla99} 
obtained maps of HCN, DCN, HNC and HDO at 2--4\arcsec~spatial resolutions over 
2--3~h integration time. 
The presence of arc-like structures offset from the nucleus is reported 
for all species but HCN, and interpreted in terms of jets of icy particles 
releasing unalterated gas contrasting with that outgassed from the nucleus. 
 
Interferometric observations of rotational lines in comet 
Hale-Bopp were also made with the Plateau de Bure interferometer 
(PdBI) of the Institut de Radio Astronomie Millim\'etrique (IRAM) 
at 1--3\arcsec~resolution. A short and preliminary account of 
these observations was given in \citet{wink99}, \citet{des99} and 
\citet{hen02}. Millimetre lines of CO, HCN, CS, HNC, CH$_3$OH, 
H$_2$S, SO, H$_2$CO were mapped \citep{wink99,boi+07}. At the same 
time, continuum maps of the dust and nucleus thermal emissions 
were obtained \citep{altenhoff}. The PdBI was also used in 
single-dish mode to detect new cometary molecules 
\citep{domi2000,cro04a,cro04b}. 
 
We present here observations of the CO $J$(1--0) (115 GHz) and 
$J$(2--1) (230 GHz) lines performed 
on 11 March, 1997 at the Plateau de Bure interferometer. Among the $\sim$20 
molecules identified in comet Hale-Bopp, and more generally in 
cometary atmospheres, carbon monoxyde CO is of particular interest: 
 
\begin{enumerate} 
\item 
This  species  is the main agent of distant cometary activity, as first 
evidenced in comet 29P/Schwassmann-Wachmann 1 \citep{sen94,cro95}. 
This was later confirmed 
in comet Hale-Bopp from its long-term monitoring, which showed the change 
from a CO-dominated to an H$_2$O-dominated activity at heliocentric distances 
$r_h$ $\sim$ 3--4 AU \citep{biv97,biver99}. 
\item 
In comets within 3~AU from the Sun, CO is, most often, the second major 
gaseous component of the coma after water. CO production rates 
relative to water are 
highly variable from comet to comet, ranging from less than 1 \% to 
\mbox{$\sim$ 20~\%} \citep[][ for a review]{irv00,domi2004}. 
That measured in comet Hale-Bopp near 
perihelion is among the highest ever observed in comets: $\sim$ 20 \% 
\citep[e.g.,][]{domi2000,disanti01}. 
\item 
 There is much debate on CO production mechanisms. Because 
CO has a low sublimation temperature, the nucleus surface is 
certainly deprived of CO ice. Therefore, CO should outgass at some 
depth inside the nucleus, possibly from pure CO ice sublimation 
and/or from amorphous water ice when crystallizing and releasing 
trapped molecules \citep[e.g.,][]{enz98,cap00,cap02}, if indeed 
pre-cometary ices condensed in amorphous form, which is somewhat 
debated \citep{mou00}. Another mechanism proposed to explain the 
CO production of comet Hale-Bopp near perihelion is the release of 
CO trapped in crystalline water ice during water ice sublimation 
\citep{cap00,cap02}. Comparing the CO jets morphology, as the 
nucleus rotates, to those of less volatile species or dust might 
provide clues to the origin of CO. \item There are several 
observational evidences that a significant part of the CO observed 
in cometary atmospheres could be produced by a distributed source. 
From in situ measurements of the local CO density in 1P/Halley 
with Giotto, \citet{ebe87} concluded that only about 1/3 of the CO 
originated from the nucleus. The spatial distribution of CO 
molecules deduced from infrared long-slit observations of comet 
Hale-Bopp led \citet{dis99,disanti01} to suggest that one-half of 
the CO was released by a distributed source when comet Hale-Bopp 
was within 1.5 AU from the Sun. The spatial resolution of the CO 
maps obtained at PdBI, approximately 1\,000 km to 1700 km radius 
on the comet depending of the line observed, is below the 
estimated radial extension of the CO distributed source of $\sim$ 
5 $\times$ 10$^{3}$ km \citep{disanti01,bro03}. Therefore, an 
important aspect of the study of the CO PdBI interferometric data 
is that independent information concerning the existence of a CO 
distributed source in Hale-Bopp coma can be possibly obtained. The 
study of the radial distribution of the CO molecules is presented 
in a separate paper \citep{Boc09}. The brightness distribution of 
both 115 GHz and 230 GHz lines can be fully explained by pure 
nuclear CO production, provided that opacity effects and 
temperature variations in the coma are taken into account 
\citep{Boc05,Boc09}. 
 
\end{enumerate} 
 
The observations of the CO $J$(1--0) and 
$J$(2--1) lines, performed both in single-dish and interferometric 
modes, are presented in Sect.~\ref{obs}. They show evidence for a rotating 
CO jet. A model simulating 
a CO spiralling jet when the nucleus is rotating is developed in 
Sect.~\ref{model}. It allowed us to compute synthetic spectra, visibilities 
and interferometric maps as a function of time, to be compared to the 
observations. Observations are analysed in Sect.\ref{jets}, and the 
model free parameters are retrieved. A discussion is given in Sect.~\ref{ccl}.

\section{Observations\label{obs}} 
 
\subsection{Description} 
\label{description} 
 
Comet C/1995 O1 (Hale-Bopp) was observed from 6 March to 22 March, 1997 with the 
Plateau de Bure interferometer of IRAM, located in the French Alps. The observations 
of the $J$(2--1) (230.538 GHz) and $J$(1-0) (115.271 
GHz) CO rotational transitions were carried out on 11 March, from 4h to 15h UT. 
On this day, comet Hale-Bopp was at the geocentric distance $\Delta=1.368$ AU and 
heliocentric distance $r_h=0.989$ AU. The weather conditions were good to excellent 
and the atmospheric seeing was $\sim$0.4 \arcsec\  for both 1.3 and 3 mm receivers. 
 
The comet was tracked using orbital elements provided by D.K. 
Yeomans (JPL, solution 55). The ephemeris was computed by P. Rocher 
(IMCCE, Observatoire de Paris) with a program which takes into 
account planetary perturbations.  The first interferometric maps 
obtained on 9 March (HCN $J$(1--0) line) revealed that both 
continuum and molecular peak intensities were offset by about 
5--6\arcsec~North in declination (Dec) from the nucleus position 
provided by the ephemeris. Observations on March 11 were made with 
the ephemeris corrected by 6\arcsec~North in Dec.

\begin{figure} 
\resizebox{\hsize}{!} 
{\includegraphics[angle=270]{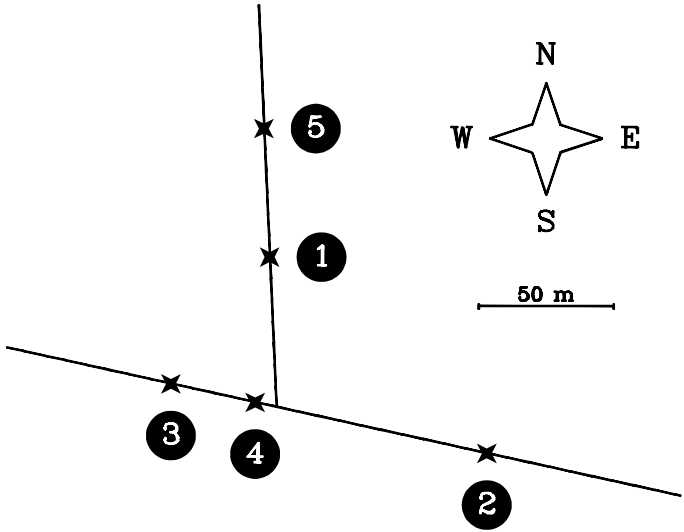}} 
\caption{The Plateau de Bure Interferometer in C1 configuration} 
\label{pdb} 
\end{figure}

The PdBI was used in the compact configuration C1 (see 
Fig.~\ref{pdb}) with five 15-m antennas providing 10 baselines 
(the spacing between two antennas) ranging from $\sim$20 to 
$\sim$150 m. In 1997, the PdBI comprised a flexible spectral 
correlator made of six independent units, providing correlated 
spectra with 64 to 256 channels spaced by 0.039 MHz to 2.5 MHz. We 
used 256 channels of 78 kHz separation for the observations of the 
230 GHz line, and 256 channels of 39 kHz separation for those of 
the 115 GHz line. The four other units were used for the continuum 
observations presented in \citet{altenhoff}. The effective 
spectral resolution is a factor of 1.3 broader than the channel 
spacing, and corresponds to $\sim$ 0.13~km~s$^{-1}$ for both 
lines. 
 
\begin{figure} 
\resizebox{\hsize}{!} {\includegraphics{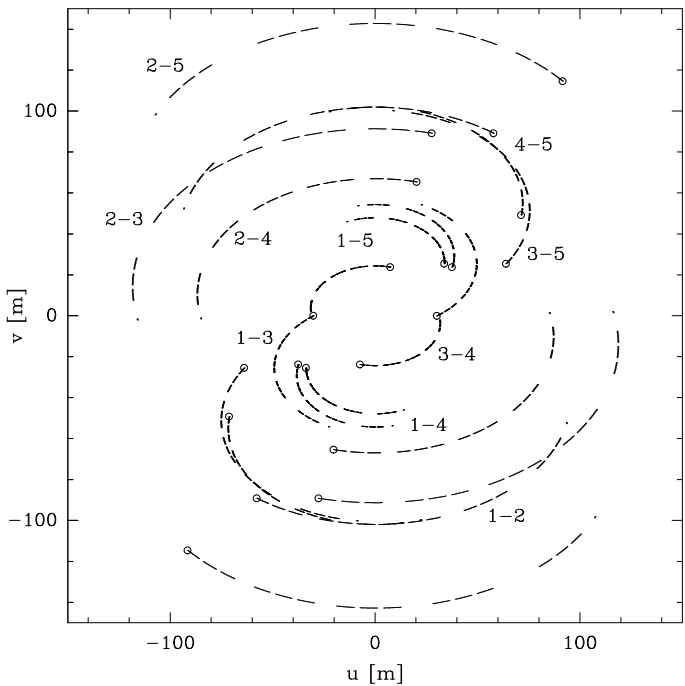}} 
\caption{{\it uv}-coverage at the PdBI on 11 March, 1997 with the 
C1 configuration. For each arc of ellipse, the white circle 
represents the $uv$-point at the beginning of the observations. 
The time evolution of the $uv$-points loci is counterclockwise. } 
\label{couv} 
\end{figure} 
 
The observing cycle was~: pointing, focusing, 4 min of 
cross-correlation on the calibrators (2200+420 BL Lac, 
MWC349 and 3C373), 2 min of autocorrelation, and 51 one minute scans 
of cross-correlation on the comet interlaced with scans on the
phase calibrator (2200+420 BL Lac) observed every 20 min. The cycle 
was completed by another 
2 min of autocorrelation on Hale-Bopp. For the autocorrelation 
observations (in this mode, the five antennas behave as five independent 
single-dish telescopes), we used position-switching (ON--OFF) with 5\arcmin{} 
offset to cancel the sky background. The spectra of the five 
antennas were then co-added. Hereafter, these autocorrelation 
observations will be denominated as ON--OFF observations. 
 
The amplitude and phase calibrator was 2200+420. MWC349 was used to determine the 
flux density of 2200+420. Bandpass calibration was made on 3C273.  Because of 
less accuracy in phase calibration after 12.5 h UT, only interferometric data 
acquired before 12.5 h UT were considered.  Calibration was done with the IRAM CLIC software, 
and the data hence derived were stored in {\it uv}-tables.  Reduction 
and cleaning of the maps were performed with the MAPPING/GILDAS software\footnote{http://www.iram.fr/IRAMFR/GILDAS/}. 
 
Concerning ON--OFF spectra, antenna temperatures $T_{\rm A}^*$ 
were converted into main beam brightness temperatures $T_{\rm mB}$ 
through  $T_{\rm mB}= T_{\rm A}^* F_{\rm eff}/B_{\rm eff}$ with 
beam efficiencies $B_{\rm eff}$ of 0.83 (115 GHz) and 0.58 (230 
GHz), and forward efficiencies $F_{\rm eff}$ of 0.93 (115 GHz) and 
0.89 (230 GHz). Flux density per beam ($S$ in Jy) is then related to antenna 
temperature through $S/T_{\rm A}^* = (2k\Omega_{\rm mB}/\lambda^2) F_{\rm eff}/B_{\rm eff} = 19.61 F_{\rm eff}/B_{\rm eff}$, where $\Omega_{\rm mB}$ is the 
main beam solid angle. 
 
For both ON--OFF and interferometric data, the uncertainties in 
flux calibration are at most 10\% and 15\% for the $J$(1--0) and 
$J$(2--1) lines, respectively. The r.m.s in phase noise ranges 
from 10 to 27\degre at 230 GHz and from 4.6 to 20\degre at 115 GHz, 
depending of the baseline.
 
The cross-correlated spectra produce, for each spectral channel, 
interferometric maps with spatial resolutions corresponding to the 
{\it uv}-coverage (Fig.~\ref{couv}). When all cross-correlation 
data are considered, the full width at half maximum (FWHM) of the 
synthesized beam is 2.00\arcsec \ $\times$ 1.38\arcsec \ with 
major axis at position angle $pa=99.56$\degre at 230 GHz, and 
3.58\arcsec \ $\times$ 2.57\arcsec \ with $pa=86.00$\degre  at 115 
GHz.  The FWHM of the primary beam of the antennas is 20.9\arcsec 
\ at 230~GHz and 41.8\arcsec \ at 115~GHz. 
 
\begin{figure*} 
{\includegraphics[bb = 110 248 481 543,angle=0,width=\textwidth, 
clip=true]{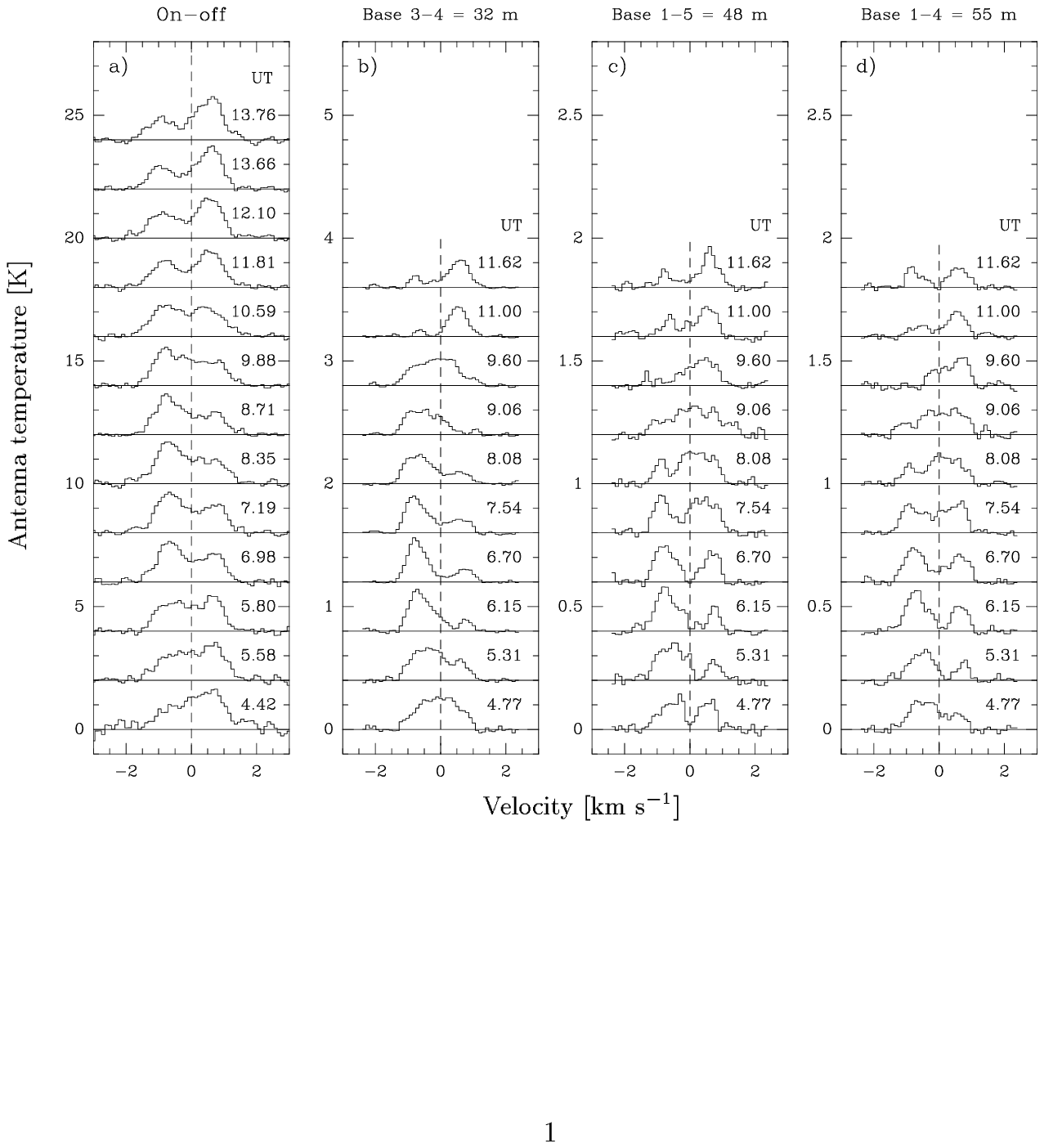}}  
\caption{CO $J$(2--1) spectra 
obtained in ON--OFF and interferometric modes as a function of 
time, given in UT hours on 11 March, 1997. The velocity scale is 
with respect to the comet rest velocity. Spectra have been shifted 
vertically according to observation time, but are scaled 
identically. {\bf a)~}ON--OFF spectra: the integration time is 2 
min (on+off). {\bf b--d)~}Interferometric spectra (visibility 
amplitude as a function of spectral channel): averages of 25 to 27 
scans of 1 min. Only data for the 3 shortest baselines are shown.} 
\label{sp-int} 
\end{figure*} 
 
\subsection{ON--OFF spectra} 
\label{sect-onoff} 
 
ON--OFF spectra of the CO $J$(2--1) line are shown in 
Fig.~\ref{sp-int}a.  The integration time is 2 minutes (on+off) 
for each spectrum.  They show a feature moving from positive to 
negative velocities, and to positive velocities again, with 
respect to the nucleus velocity frame. In other words, a jet-like 
CO gas feature, which velocity vector with respect to Earth 
rotated during the course of the observations, is observed. This 
CO gas feature contributes up to 28\% of the total line area. 
 
\begin{figure} 
\resizebox{\hsize}{!} 
{\includegraphics[angle=270]{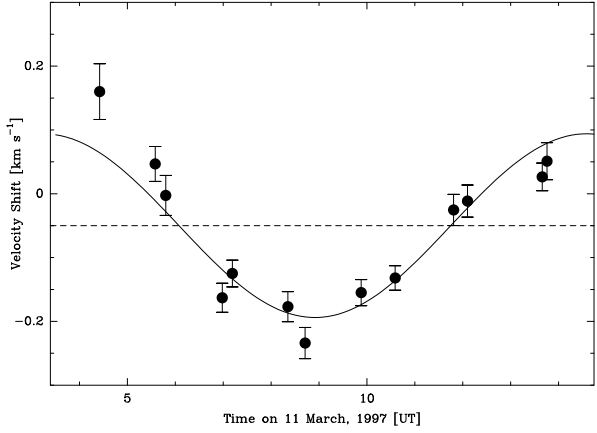}} 
\caption{Time 
evolution of the velocity shift of CO $J$(2--1) ON--OFF spectra 
 shown in Fig.~\ref{sp-int}a. 
The plotted curve is the least-squares sinusoid fit to the data. 
It has a fixed period of 11.35\,h, an amplitude of 
0.29$\pm$0.03\,km\,s$^{-1}$, and a velocity centre $v_{\rm 0}=-0.05\pm$0.01 km\,s$^{-1}$ (dotted line).} 
\label{evol-velo} 
\end{figure}

The synodic rotation period of comet Hale-Bopp in February--April 1997 has 
been deduced from studies of the dust shells \citep{sarmecanic97,farnham99,ortiz99}. 
The most accurate value $P$ = 11.31 $\pm$ 0.01 hr measured by \citet{farnham99} 
is in agreement with a slightly larger sidereal rotation period $P$ = 11.34 $\pm$ 
0.02 hr \citep{licandro98,jorda99}. Figure~\ref{evol-velo} plots the evolution of the 
line velocity shift (the spectrum first order momentum) with time.  The points 
follow a sinusoidal curve which period corresponds to the comet's nucleus 
rotation (taken to be equal to 11.35 hr throughout this paper). 
The sinusoid determined from a least-squares fit with fixed period $P$ = 11.35~h has a mean level of $v_{\rm 0}=-0.05\pm0.01$~km\,s$^{-1}$ and its amplitude is $\mathcal{A}$ = 0.29$\pm$0.03~km\,s$^{-1}$ (Fig.~\ref{evol-velo}). 
The line area does not show significant variation with time, with a mean value of 
4.22~$\pm$~0.03~K\,km\,s$^{-1}$ in main beam brightness temperature scale $T_{\rm mB}$ 
(i.e., 82.8 Jy km~s$^{-1}$ in flux density scale). 
Fluctuations of $\sim$10\% at most are observed (with a standard deviation of 5\%), 
which are not correlated with the velocity shift variations. The velocity shift curve 
obtained for the $J$(1--0) line is much more noisy (error bars $\sim$0.2 km\,s$^{-1}$ on 
the individual spectra), but is similar to the $J$(2--1) velocity shift curve: 
a least-squares sinusoid fit with $P$ fixed to 11.35~h leads to 
$\mathcal{A}$ = 0.39$\pm$0.16~km\,s$^{-1}$, $v_{\rm 0}=-0.05\pm0.05$~km\,s$^{-1}$  \citep{hen03}. 
Adding all spectra, the mean velocity shift of the $J$(1--0) line is 
$-$0.09~$\pm$~0.05~km\,s$^{-1}$, in agreement 
with that of the $J$(2--1) line ($-$0.083 $\pm$ 0.007~km\,s$^{-1}$). The line area 
of the $J$(1--0) line is 0.552~$\pm$~0.022~K\,km\,s$^{-1}$ in the $T_{\rm mB}$ scale (i.e., 
10.8 Jy km~s$^{-1}$).

From the spin axis orientation and the equatorial coordinates of the comet, 
it is possible to derive the angle \thw (aspect angle) between the spin axis 
and the line of sight, and the North pole position 
angle \paw, defined from North to East. 
The different spin orientations published in the literature are listed in 
Table~\ref{polpos}. 
 
\begin{table} 
\caption{Spin axis orientation. 
Columns 1 and 2 are the equatorial coordinates found in the literature. 
Columns 3 and 4 are the corresponding position angle (\paw) and aspect angle (\thw) 
on 11 March, 1997. } 
\label{polpos} 
\begin{tabular}{llllll} 
\hline 
\multicolumn{1}{c}{$\alpha_{\omega}$} & 
\multicolumn{1}{c}{$\delta_{\omega}$} & 
\multicolumn{1}{c}{$pa_{\omega}$} & 
\multicolumn{1}{c}{$\theta_{\omega}$} & 
\multicolumn{1}{c}{Epoch} & 
\multicolumn{1}{c}{Ref.$^{a}$} \\ 
\hline 
\phantom{0}30\degre & \phantom{-}45\degre & \phantom{0}66\degre & 143\degre & May--Nov 1996 & [1] \\ 
170\degre & -40\degre & 272\degre & \phantom{00}9\degre & May--Nov 1996 & [2] \\ 
240\degre & -56\degre & 224\degre & \phantom{0}53\degre & May--Nov 1996 & [2] \\ 
275\degre & -50\degre & 217\degre & \phantom{0}74\degre & Mar--Nov 1996 & [3] \\ 
320\degre & -60\degre & 189\degre & \phantom{0}78\degre & Sep 1995--Jan 1998 & [4] \\ 
290\degre & -40\degre & 215\degre & \phantom{0}88\degre & Sep 1996--May 1997 & [5] \\ 
290\degre & -60\degre & 203\degre & \phantom{0}72\degre & Feb 18, 1997 & [6] \\ 
255\degre & -60\degre & 216\degre & \phantom{0}59\degre & Feb 1997 & [2] \\ 
275\degre & -57\degre & 211\degre & \phantom{0}68\degre & Feb--Mar 1997 & [7] \\ 
276\degre & -54\degre & 213\degre & \phantom{0}71\degre & Apr 1996--May 1997 & [8] \\ 
\hline 
\end{tabular} 
\begin{list}{}{} 
\item[$^{a}$] [1]~:~\citet{sekanina97}, [2]~:~\citet{sekanina99}, [3]~:~\citet{licandro99}, [4]~:~\citet{biver98}, [5]~:~\citet{metchev98}, [6]~:~\citet{vas99a}, 
[7]~:~\citet{jorda99}, [8]~:~\citet{sch04}. 
\end{list} 
\end{table} 
 
Adopting the spin orientation derived by \citet{jorda99} and \citet{sch04}, 
the comet spin axis was then only 20\degre\ far from the plane of the sky. 
In such a geometrical configuration, a polar CO jet would lead to an almost 
constant velocity shift. A jet close to the equator can explain a 
velocity shift following a sinusoid centred around $v_{\rm 0}$ $\sim$ 0 km s$^{-1}$. 
Both this sinusoidal curve and the constant CO line area show that 
the amount of CO gas released in this jet did not vary 
during nucleus rotation. 
Given the Sun direction (phase angle of 46\degre, $pa$ = 160\degre), this 
near-equatorial CO jet was active 
night and day at about the same extent. 
 
From the line areas of the $J$(1--0) and $J$(2--1) ON--OFF 
profiles, we derive a CO production rate $Q_{\rm CO}$ =  2.1 
$\times$10$^{30}$ s$^{-1}$. Here, we have assumed a Haser parent 
molecule distribution for CO, and run our excitation model 
(Sect.~\ref{model}) with a kinetic temperature $T$ of 120 K which 
agrees with temperature determinations pertaining to the 
10\,000--20\,000 km (radius) coma region sampled by the primary 
beam of PdBI \citep{biver99,disanti01}. Using an extended 
production for CO consistent with the IR observations does not 
significantly affect the inferred $Q_{\rm CO}$. \citet{disanti01} 
inferred a total CO production rate (nuclear+distributed) fully 
consistent with our value. In the following sections, we will assume 
$T$ = 120 K and a  
total $Q_{\rm CO}$ of 2 $\times$10$^{30}$ s$^{-1}$.

\subsection{Interferometric data} 
\label{inter-data} 
 
\begin{figure} 
{\includegraphics[width=9.5cm,angle=270]{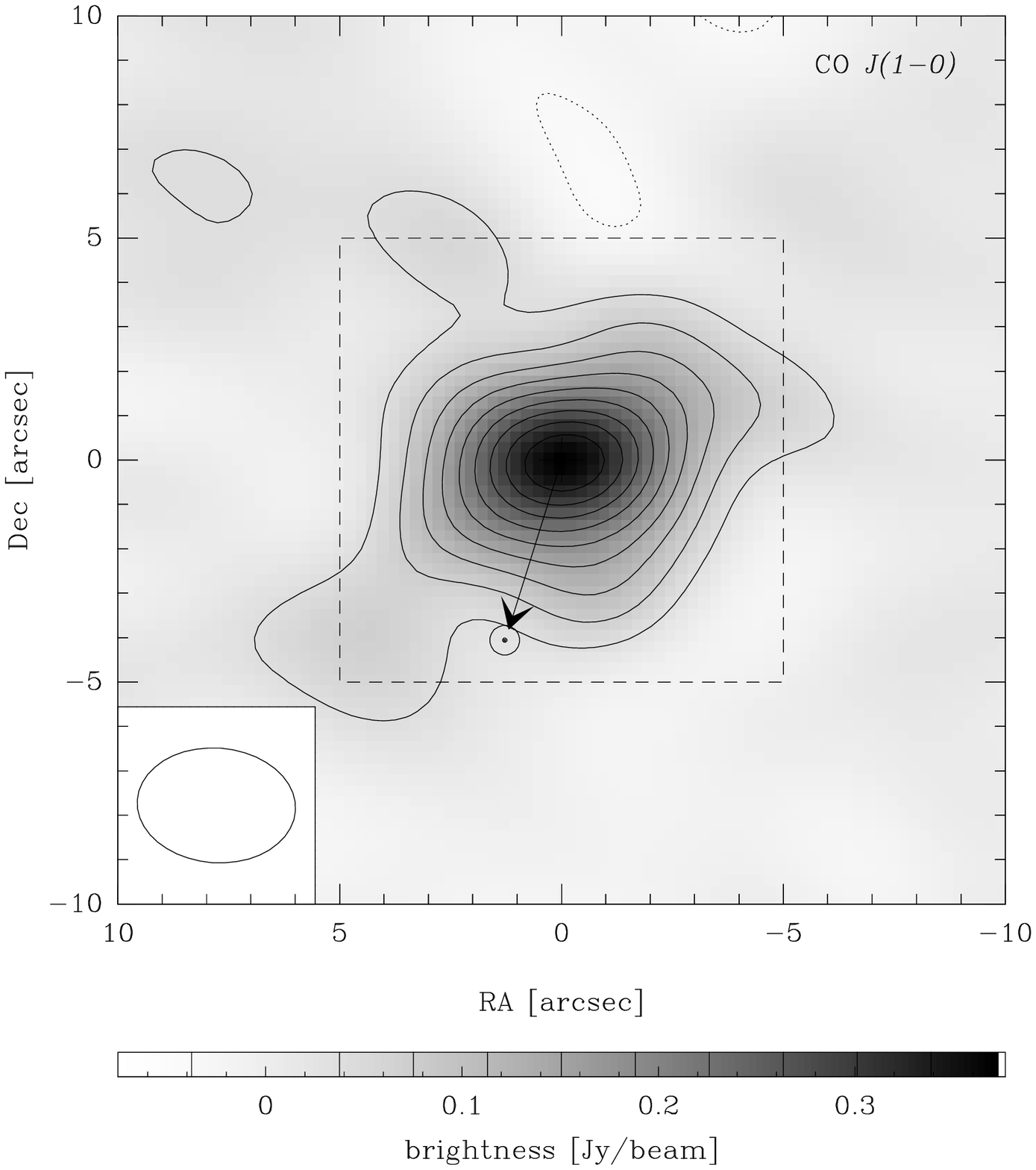}} 
\vspace{+0.8cm} 
{\includegraphics[width=9.5cm,angle=270]{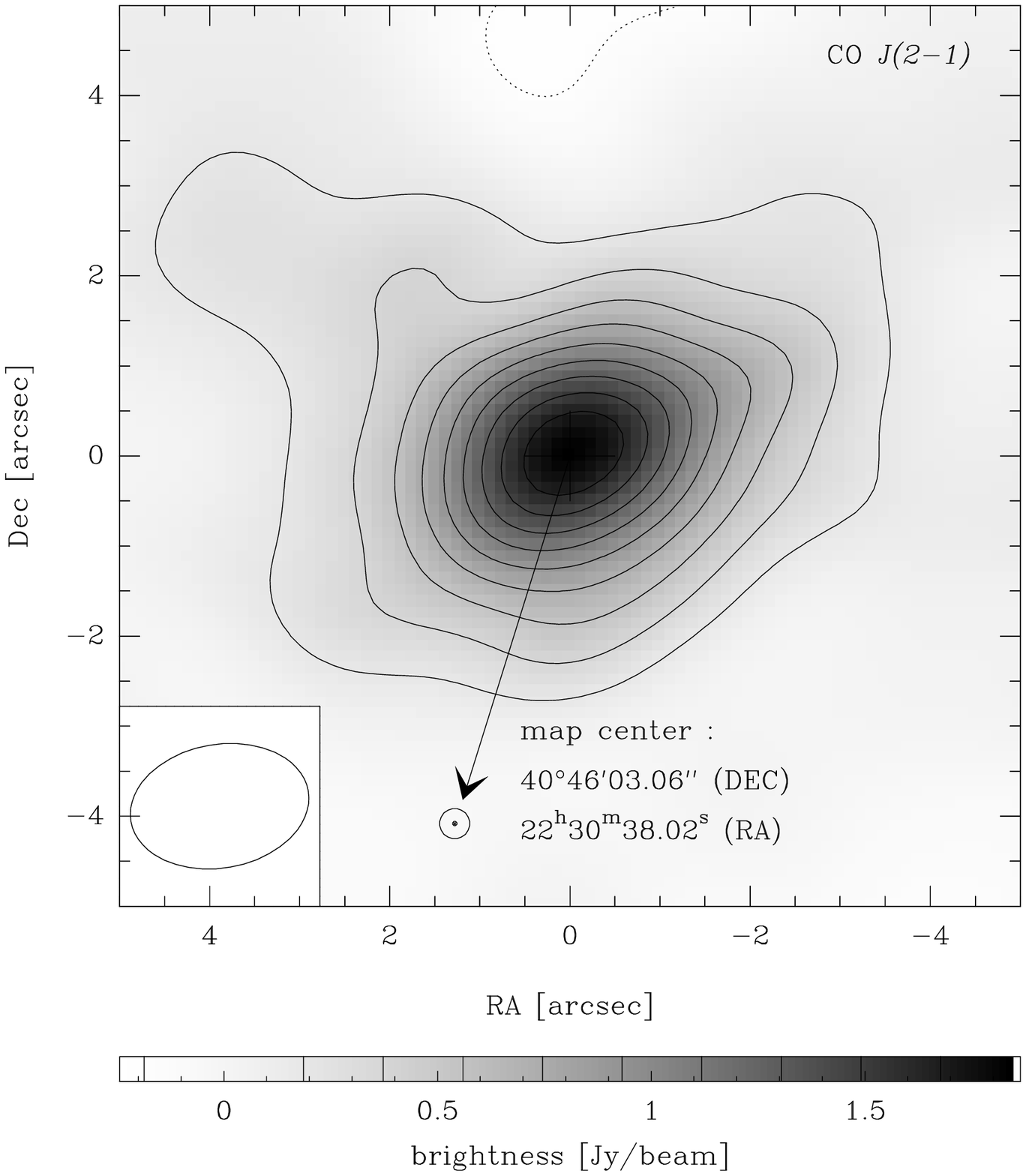}} 
\caption{CO $J$(1--0) (top) and $J$(2--1) (bottom) line integrated maps observed on 11 March, 1997 (all 
data). 25 spectral channels have been averaged. The synthesized beam is in the lower left. The 
dashed square on the $J$(1--0) map corresponds to the size of the $J$(2--1) map. 
For the CO $J$(1--0) line, contours are 0.037 Jy/beam, 
and the r.m.s. is 0.018 Jy/beam. For the $J$(2--1) line, contours are 0.186 Jy/beam and the r.m.s. is 0.066 Jy/beam. 
Iso-contours are successive multiples of 10\% of the maximum 
intensity, at 10 to 100 \% of the maximum intensity. The solar direction is 
indicated. The original images have been shifted so that the maximum brightness peaks at the centre of the maps. In units of line area, the intensities at   
maximum brightness are 0.93$\pm$0.04 and 4.65$\pm$0.15 Jy km s$^{-1}$ for the  
 $J$(1--0) and $J$(2--1) lines, respectively. } 
\label{co21-1} 
\end{figure} 
 
We present in Fig.~\ref{co21-1} the line integrated maps of CO 
$J$(2--1) and $J$(1--0).  8~hours of observations were averaged 
(from 4.5 h UT to 12.5 h UT) and 25 velocity channels (12 on both sides  
the central channel corresponding to the nucleus 
velocity) were co-added. An asymmetrical shape, which is not aligned with the 
elliptical clean beam (the synthesized interferometer beam), is 
observed and related to the anisotropy of the gas emission. 

\begin{figure*} 
\centering 
{\includegraphics[angle=270,width=17cm]{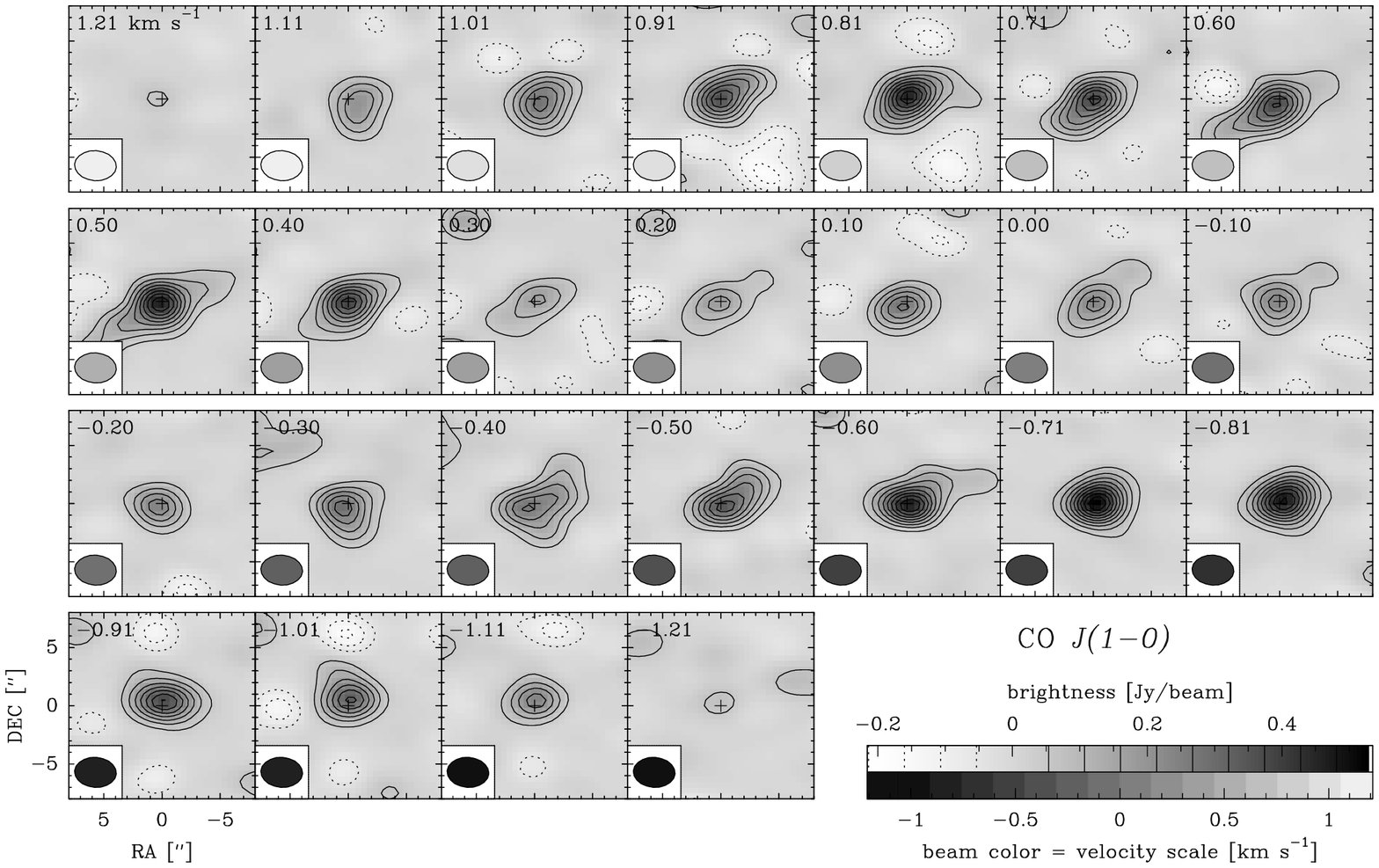}} 
\vspace{+0.8cm} 
{\includegraphics[angle=270,width=17cm]{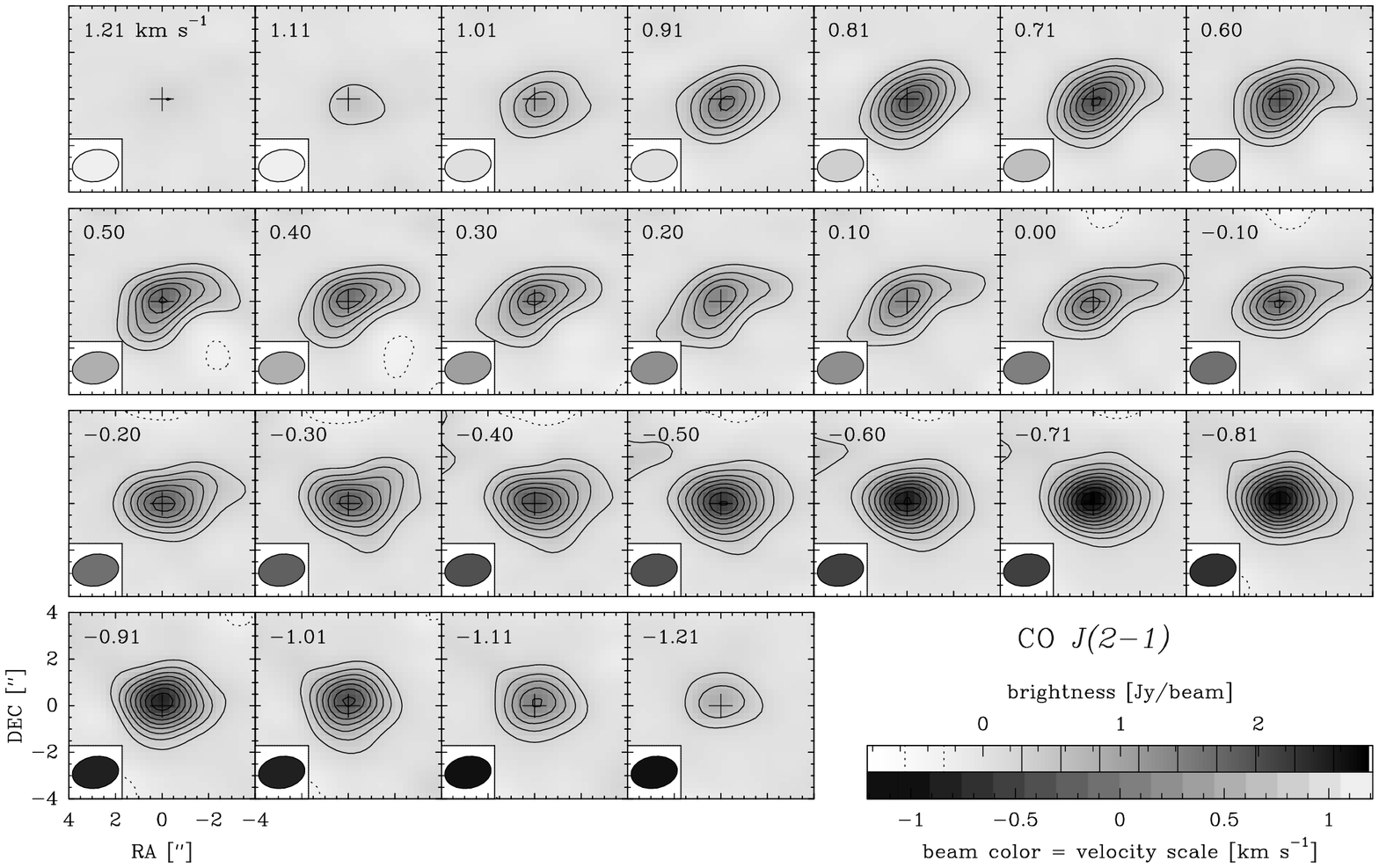}} 
\caption{CO maps as a function of spectral channel on 11 March, 1997 (all 
data averaged). RA and Dec positions are with respect to the mean photometric 
centre $C_{\rm m}$ determined from the whole data set. The 
velocity (with respect to the comet rest velocity) of the spectral channels is indicated in the 
top left corner of the maps. The synthesized beam is in the lower left, and is coloured 
according to the velocity value. $J$(1--0) line (top): contour interval is 0.053 Jy/beam 
and  the r.m.s. is 0.046 Jy/beam. $J$(2--1) line (bottom): contour interval 
is 0.283 Jy/beam and the 
r.m.s. is 0.15 Jy/beam. Contours correspond to multiples of 10\% the peak flux density measured on channels at Doppler velocities of $-$0.71 and $-$0.81 km 
s$^{-1}$.} 
\label{co21-25} 
\end{figure*} 

\begin{figure*} 
\includegraphics[angle=270,width=17cm]{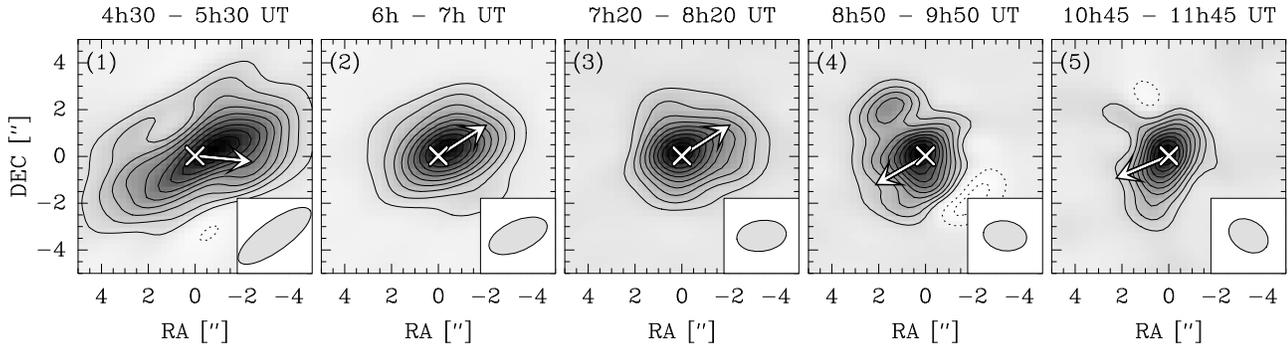} 
\caption{Individual maps of CO $J$(2--1) for data subsets of 1~h. 
Iso-contours are successive multiples of 10\% of the maximum 
intensity, at 10 to 100 \% of the maximum intensity. 
For each map labelled (i) on the top left corners, a crosse locates the mean photometric 
centre $C_{\rm m}$ determined from the whole data set. The arrow represents the 
direction of the individual photometric 
centre $C_{\rm i}$ with respect to $C_{\rm m}$.  $C_{\rm i}$ is evaluated by fitting a 
2-D Gaussian of 
adjustable width. The beam shape is shown in the bottom right corner. 
\label{evolt5}} 
\end{figure*} 
 
\begin{figure} 
\resizebox{\hsize}{!}{ 
\includegraphics[angle=270]{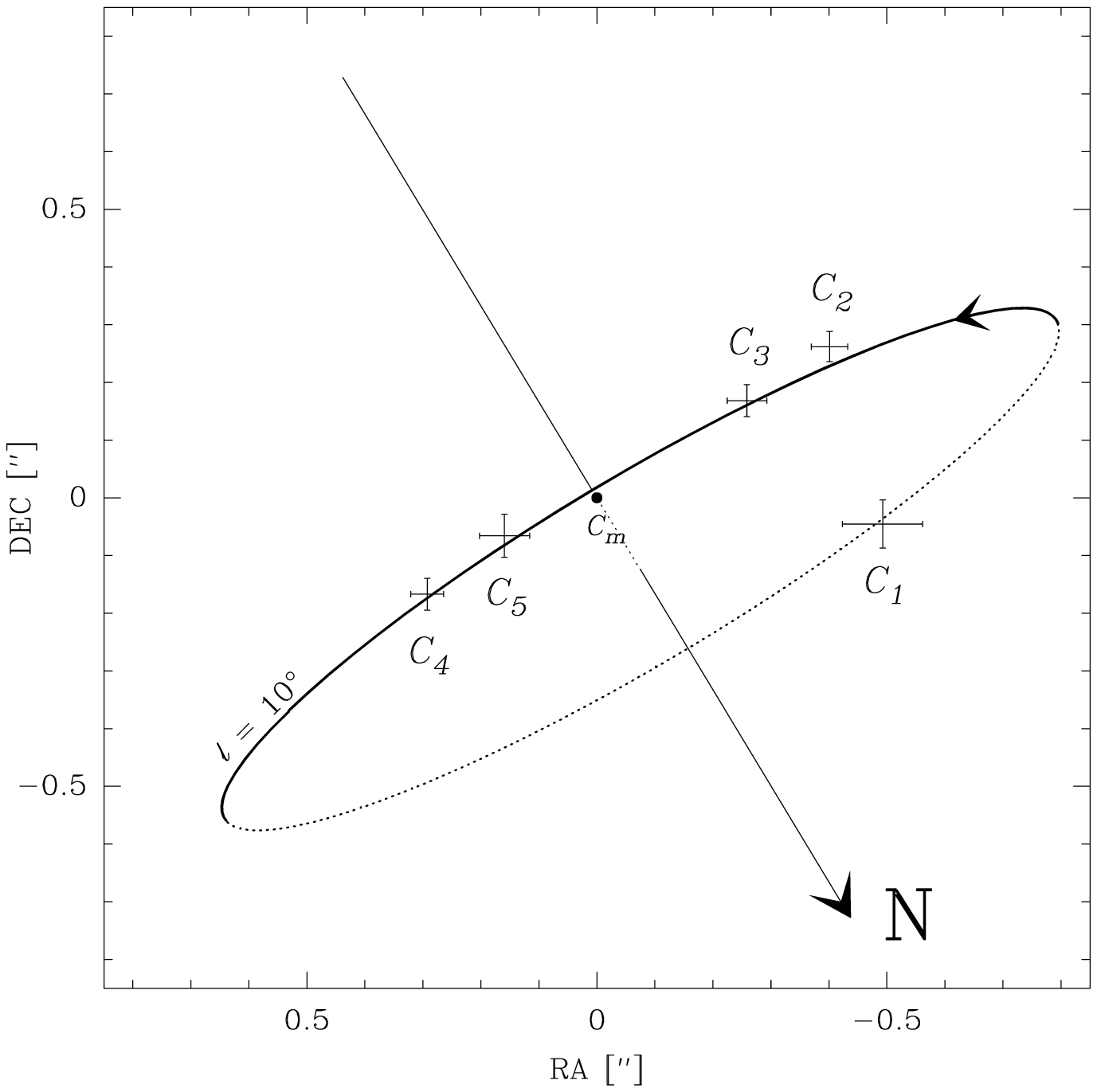}} 
\caption{Time evolution of the photometric centres. 
Photometric centres $C_{\rm i}$, with $i$ refering to the maps shown in 
Fig.~\ref{evolt5}, 
are given with their errorbars. The ellipse drawn is a least-squares fit to the data. 
It corresponds to a parallel of latitude 10\degre North of a 
sphere, with spin axis aspect angle~=~80\degre and position angle~=~211\degre. 
\label{evolt5-center}} 
\end{figure}

In the line integrated interferometric map of CO $J$(2--1) 
(Fig.~\ref{co21-1}), the position of the peak brightness  ( 
$C_{\rm m}$) is at RA = 22h\,30m\,38.02s and Dec = 
40\degre\,46\arcmin\,3.1\arcsec~(with an astrometric precision of 
0.07\arcsec) in apparent geocentric coordinates given for 7.00 h UT. 
 The peak position of the CO $J$(1--0) brightness (RA = 
22h\,29m\,38.46s and Dec = 40\degre\,41\arcmin\,10.1\arcsec~at 4.00 h 
UT) is consistent with that of $J$(2--1), taking into account the 
comet motion from 4 to 7 h UT \citep[see][]{boi+07}. The peak 
position of the continuum emission at 230 GHz observed 
simultaneously also almost coincides (0.2\arcsec~offset) with the 
CO  $J$(2--1) peak \citep{altenhoff,boi+07}.   
Using orbital elements based on optical astrometric positions  
from April 1996 to August 2005 (JPL solution 220),  
the offset between the CO peak and the ephemeris is 
+2.9\arcsec\, in Dec and +0.4\arcsec\, in RA. So, positions of the 
CO and radio continuum brightness peaks differ by typically  
+3\arcsec\ in Dec from optical astrometric positions. A bright 
dusty jet was identified southward in the optical images of comet 
Hale-Bopp near perhelion \citep[e.g.,][]{jorda99}. As shown by 
\cite{boi+07}, the optical astrometric positions were more 
affected by dusty jets than the radio positions. \citet{boi+07} showed  
that the astrometric positions provided by the IRAM continuum radio maps 
provides an orbit which does not require the existence of 
non-gravitational forces acting on the Hale-Bopp nucleus, in 
contrast to those derived from only optical positions, thereby 
solving a contentious issue. In conclusion, there 
is no substantial offset between the nucleus position and the mean 
photometric centre of CO emission.

The $J$(1--0) and $J$(2--1) spectral channel maps 
(Fig.~\ref{co21-25}) were obtained with the same procedure.  The 
peak brightness on the blue channels is stronger than that on the 
red ones. This indicates more emission toward the Earth and is 
related to the jet seen in ON--OFF spectra. Indeed, the 
interferometric observations covered only 2/3 of the nucleus 
rotation period, when the jet was, most of the time, facing the 
Earth (Fig.~\ref{evol-velo}). The spectral maps show that the CO 
coma structure is complex. The interpretation of the brightness 
distribution on these maps is not straightforward, since the 
signal is here averaged over the entire period of observation and 
the CO coma is rotating. The most central channels are sensitive 
to molecules expanding along directions close to the plane of the 
sky. They show coma structures towards North-West and South-East 
quadrants (roughly along a direction perpendicular to the 
projected rotation axis, see the 230 GHz maps in 
Fig.~\ref{co21-25}). These structures may trace the jet at the 
time it was near the plane of the sky. Channels at high negative 
velocities (--0.6 to --0.9 km s$^{-1}$) show a much brighter and 
strongly peaked intensity distribution because the jet is here 
facing the Earth. 
 
\begin{figure*} 
\includegraphics[angle=270,width=17cm]{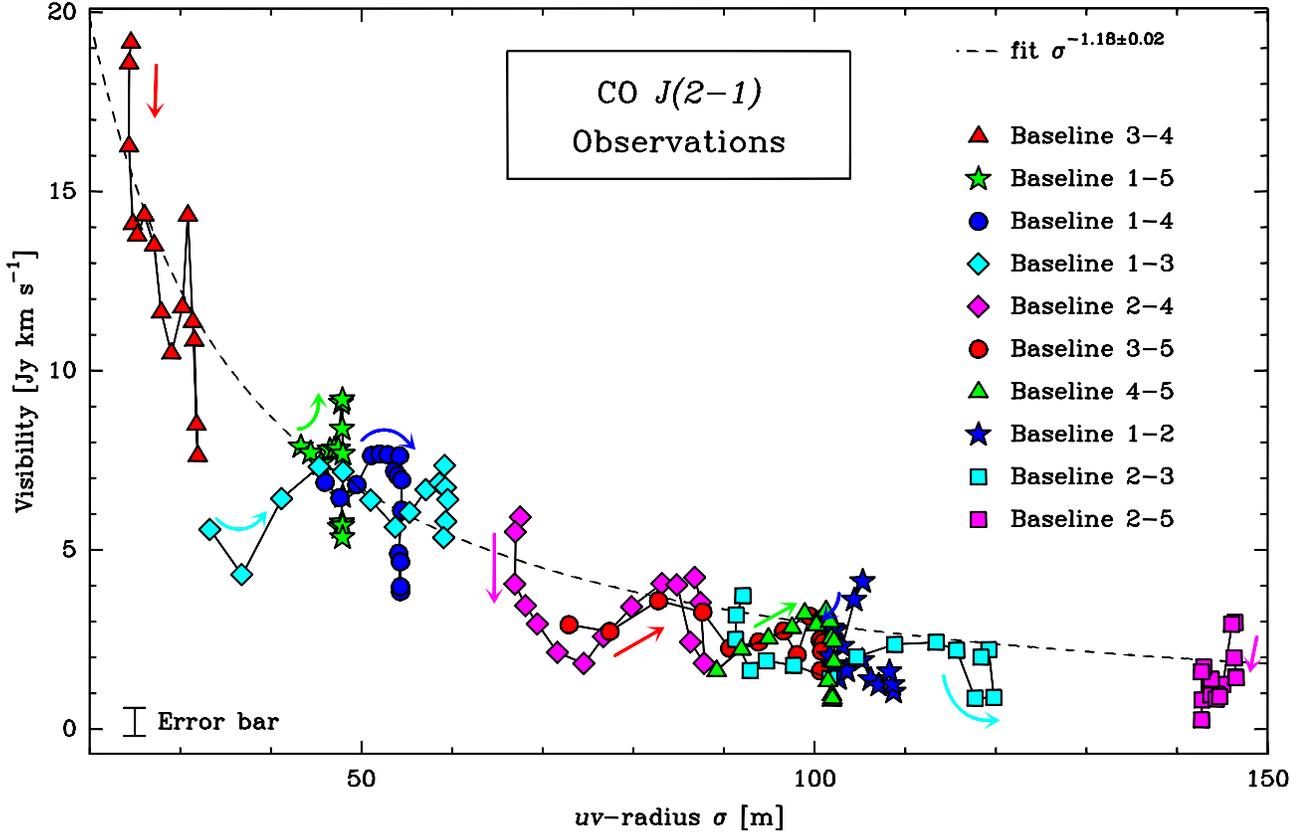} 
\caption{ 
Time evolution of the visibility amplitudes with respect to the 
$uv$-radius $\sigma$. Different symbols are used for the different baselines. 
For some baselines, the arrow shows the direction of time evolution of the 
$uv$-radius. The uncertainties on the data points due to thermal and 
phase noise range from 0.29 to 0.65 Jy km s$^{-1}$. Phase noise affects the 
uncertainties by 2\% (short baselines) to 10\% (long baselines). 
The mean error bar ($\pm$ 0.40\,Jy km s$^{-1}$) is quoted 
on the figure. The dashed curve is a least-squares fit of a power law to the 
data. } 
\label{visi-obs} 
\end{figure*} 
 
In order to investigate whether there is temporal evidence for the 
rotating jet in the images of the CO emission, we have combined 
the data into five parts of about 1 hour each. Resulting maps are 
presented in Fig.~\ref{evolt5}.  Because of Earth rotation, the 
beam shape rotates with time from map to map and changes dimension 
(see next paragraph). This prevents a detailed study of the 
rotating jets directly from the maps and, as explained later, 
another approach will be used. However, an interesting feature is 
observed.  From the observations averaged over the whole day, we 
have derived the mean photometric centre of CO emission, $C_{\rm 
m}$. For each map $i$, we can also derive the photometric centre, 
$C_{\rm i}$, and the vector $\vec{J_{\rm 
i}}=\overrightarrow{C_{\rm m} C_{\rm i}}$, as shown in 
Fig.~\ref{evolt5}. The time evolution of $\vec{J_{\rm i}}$ is 
presented in Fig.~\ref{evolt5-center}. We observe that it moves 
counterclockwise (disregarding $C_{\rm 5}$) along an ellipse which 
long axis is perpendicular to the spin axis direction. Such a 
displacement is that expected in presence of a CO rotating jet. 
Provided $C_{\rm m}$ coincides with the nucleus position, 
$\vec{J_{\rm i}}$ reflects the jet direction on map $i$. For a 
spherical nucleus and constant jet activity, the $C_{\rm i}$'s 
locus should be then an ellipse which long axis position angle is 
perpendicular to the spin axis, the other characteristics of the 
ellipse (axis lengths, centroid) being related to the amount of CO 
gas inside the jet, as well as to its latitude on the nucleus 
surface. A least-squares fit of the photometric centres leads to 
an ellipse (Fig.~\ref{evolt5-center}) that corresponds to a spin 
axis with position angle \paw=\,211\degre\ and aspect angle 
\thw=\,79\degre, in good agreement with most of the published 
values (Table~\ref{polpos}). The ellipse dimensions and position 
inferred from the fit do not provide direct quantitative 
information on the jet relative strength and latitude because the 
nucleus position may be off by a fraction of an arcsec with 
respect to $C_{\rm m}$. However, the significant displacement of 
the photometric centre during nucleus rotation excludes a 
high-latitude jet, in agreement with the conclusion obtained from 
the ON--OFF spectra. The small offset between $C_{\rm m}$ and the 
nucleus position (as determined from the peak of the continuum 
emission) is also consistent with a low-latitude jet.

For a deeper study of the interferometric data, we have decided to work on complex 
visibilities in the {\it uv}-plane. For readers not familiar with interferometry, 
let us explain briefly what this means and how maps are obtained. An interferometer 
measures the Fourier transform (FT) of the source brightness distribution on the 
sky. The complex visibilities $\mathcal{V}$$(u,v)$ sample the FT at points $(u,v)$ 
in the Fourier plane, also called the {\it uv}-plane. These points are the 
projections of the baselines onto the plane of the sky and define the {\it uv}-coverage 
of the observations (Fig.~\ref{couv}).  As the Earth rotates, the locus of the 
points $(u,v)$ produced by one baseline is an arc of ellipse.  Therefore, the 
$uv$-radius $\sigma = \sqrt{u^2+v^2}$ changes with time (except if 
the source observed is circumpolar, because the locus is then a circle).  The 
longer we observe, the longer are these arcs, the larger is the {\it uv}-plane 
coverage \citep[see][ chap.~4,~\S4.2]{thompson}.  (In fact, the {\it uv}-coverage 
produced by a couple of antennas comprises two arcs of ellipse symmetrical with 
respect to the centre $(u,v)=(0,0)$;  this is because the source brightness 
distribution is a real function, so its FT verifies $\mathcal{V}(u,v) = 
\mathcal{V}(-u,-v)$.)  In order to make a map, one has first to compute the inverse 
Fourier transform of the sampled signal.  In a second time, this {\it dirty map} 
is deconvolved from the {\it dirty beam}, which is the FT of the {\it uv}-coverage. 
Because the {\it uv}-plane is not regularly covered, interpolations are made when performing the FT.  In addition, when the {\it uv}-coverage is highly anisotropic, the dirty beam presents intense 
sidelobes, which might not be properly accounted for in the deconvolution step. 
This might result in the apparition of artefacts.  The 
anisotropic {\it uv}-coverage also results in an elliptical clean beam.

Since the individual baselines have different directions and 
lengths, they probe different scales and regions of the coma. So, 
visibilities have to be studied for each baseline separately.  We 
plot in Fig.~\ref{visi-obs} the time evolution of the visibility 
amplitude $\bar{\mathcal{V}}$ of the CO $J$(2--1) line with 
respect to the {\it uv}-radius $\sigma$. The visibilities have 
been integrated over velocity and have units of line area. Let us 
assume that the line is optically thin and its excitation does not 
vary within the field of view.  For an isotropic coma described by 
a parent molecule distribution, the visibility curve would follow 
$\bar{\mathcal{V}}(\sigma) \propto \sigma^{-1}$, provided the 
photodissociation scale length is large compared to the field of 
view, which is the case here \citep{Boc09}. We observe in 
Fig.~\ref{visi-obs} some modulations with respect to the mean 
evolution (in $\sigma^{-1.18}$) that are not due to noise. They 
cannot be due to variations of the activity of the comet since the 
area of the line, observed in ON--OFF mode, is roughly constant 
with time. Furthermore, modulations do not present the same 
behaviour from one baseline to another.

\begin{figure} 
\resizebox{\hsize}{!} 
{\includegraphics{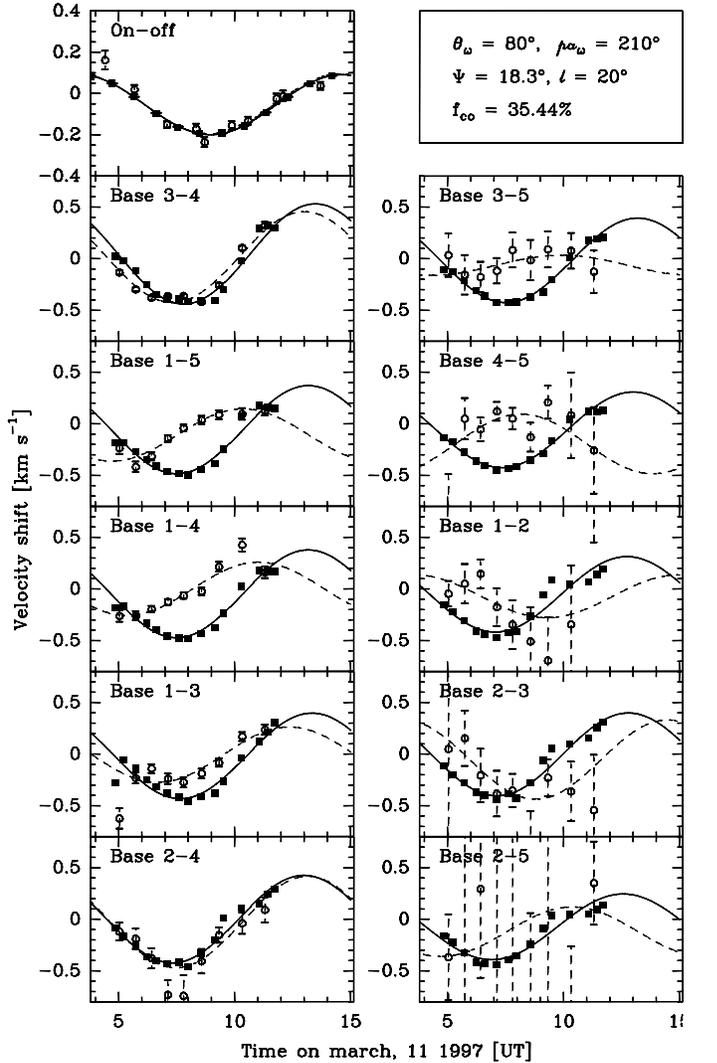}}
\caption{Time evolution of the CO $J$(2--1) velocity shifts and fitted 
sinusoids. Baselines are indicated in the top left corners. 
Observations are shown with empty circles with error bars, and 
dashed curves for the sinusoids. Model results (plain squares and plain curves) 
are for parameter set (3) of Table~\ref{selected-couples} 
with \paw~=~210\degre 
(\thw~=~80\degre, $\Psi$~=~18.3\degre, \lat~=~20\degre and \fco~=~35.5\%).} 
 
\label{evol-velo-int-mod} 
\end{figure}

We presented in Fig.~\ref{sp-int}b--d spectra issued from the 
three shortest baselines of the interferometer.  As on the ON--OFF 
spectra (Fig.~\ref{sp-int}a), we can see spectral features moving 
from red to blue velocities. Figure~\ref{evol-velo-int-mod} 
presents the time evolution of the interferometric velocity 
shifts.  At least for the five shortest baselines, they can be 
fitted by sinusoids of period equal to the nucleus rotation 
period. We observe that these curves are not in phase. As the 
baselines do not have the same length, nor the same orientation, 
this phase difference may suggest that the CO jet is spiralling 
and detected at different times with the various baselines. A 
straight jet would produce velocity shift curves in phase. 
 
The next section presents a model with a rotating gas jet aimed at 
the interpretation of these data.

\section{Model\label{model}} 
 
The model is a static 3-dimensional model simulating a spiralling gas jet. 
It computes {\it uv}-tables (i.e., visibilities) corresponding to 
the observational 
conditions of the data, that is with the same 
{\it uv}-coverage. The time evolution of the coma is simulated 
by computing a serie of successive static {\it uv}-tables. 
 
The CO coma is modelled as a combination 
of an isotropic outgassing and a gas jet~defined by: 
\begin{itemize} 
\item its half-width $\Psi$; 
\item its latitude $\ell$ on the nucleus, assumed to be spherical; 
\item the fraction \fco of the CO released within the jet. 
\end{itemize} 
The model parameters are summarized in Table~\ref{param}.

The coordinate frame $(Oxyz)$ 
used in the calculations has its origin at the comet nucleus with the $z$-axis 
along the line of sight opposite to the Earth, and the $x$ and $y$-axes 
pointing East 
and North in the plane of the sky. $r_0$ is the nucleus radius. 
The jet direction at the nucleus surface 
$\vec{J}_0$ is $(\theta_0,\phi_0)_t$ in the $(Oxyz)$ frame at time $t$ 
(usual spherical coordinates) 
and is moving with time due to the nucleus rotation. 
Let us define $\vec{R}_{\omega}$ the rotation matrix for a lapse $\Delta t$, so that 
we have $(\theta_0,\phi_0)_{t+\Delta t} = 
\vec{R}_{\omega} \cdot (\theta_0,\phi_0)_{t}$. The jet direction at distance 
$r=r_0+v_{\rm exp}\Delta t$ in the coma  is $\vec{J}_r = 
\vec{R}_{\omega}^{-1} \cdot \vec{J}_0$, where $v_{\rm exp}$ is the gas expansion 
velocity. 
 
We assume a Haser-like parent molecule distribution for CO. 
Indeed, as commented in Sect.~\ref{intro}, though infrared 
observations suggest that part of the CO in Hale-Bopp coma is 
originating from a distributed source \citep{disanti01,bro03}, the 
present observations do not require CO to be extended 
\citep{Boc09}. The local density at $(r,\theta,\phi)$ direction is 
then given by: 
 
\begin{equation} 
n_{\mathrm{CO}}(r,\theta,\phi) = \frac{Q(r,\theta,\phi)}{4\pi r^2 v_{\rm exp}} 
\exp\left(-\frac{(r-r_0)}{L_\mathrm{CO}}\right), 
\label{haser_mere} 
\end{equation} 
 
\noindent 
where 
 
\begin{equation} 
Q(r,\theta,\phi) = Q_{\rm iso} + 4 \pi\, Q_{\rm jet}\, \mathcal{G}(r,\theta,\phi). 
\end{equation} 
 
\noindent 
$Q_{\rm iso}$ and $Q_{\rm jet}$ are the total CO production rates 
due to the isotropic contribution and within the jet, respectively. 
For the sake of simplicity, the total CO production rate 
$Q_{\rm CO}=Q_{\rm iso}+Q_{\rm jet}$ is fixed, and taken to be equal to the 
2 $\times$ 10$^{30}$ s$^{-1}$ (\citet{biver99}, see Sect.~\ref{sect-onoff}).  \fco$= 
\frac{Q_{\rm jet}}{Q_{\rm CO}}$ is a free parameter. 
The function $\mathcal{G}(r,\theta,\phi)$ describes the jet pattern and is a normalized 
Gaussian $(\int\mathcal{G}(\Omega) d\Omega =1)$ of half width $\Psi$ centred on $\vec{J}_r$. 
$L_{\mathrm{CO}}=v_{\rm exp}/\beta_{\mathrm{CO}}$ is the photodissociative 
scalelength, where $\beta_\mathrm{CO}$ is the CO photodissociation rate. 
 
The code uses $N = 47$ $(Oxy)$ grids, each of them 
sampling one channel of the spectrum centred on 
$v_{\rm i}=\left(\frac{N+1}{2}-i\right)\,\delta v$, in the nucleus velocity frame, 
with $\delta v = 0.10$ km\,s$^{-1}$. The $(Oxy)$ grids are 
100\arcsec $\times$100\arcsec\ long \footnote{this is much larger than the 
primary beam of the antennas, but 
this was necessary in order to have a good resolution in the Fourier plane.}. 
They are divided into $256\times 256$ cells which dimensions are 
$(\delta x,\,\delta y)$ ($\delta x = \delta y = 0.39$ \arcsec). 

In the optically thin case, the brightness distribution in the plane of the 
sky~ [W m$^{-2}$ sr$^{-1}$], 
when selecting only molecules contributing to channel $i$, is given by: 
\begin{equation}\label{F-distri} 
F_i(x,y) = \frac{{\cal N}_i(x,y)}{\delta x\,\delta y} h \nu A_{ul} \frac{1}{4 \pi \Delta ^2}, 
\end{equation} 
where $\nu$ is the line frequency, and $A_{ul}$ is the Einstein 
coefficient for spontaneous emission. $\delta x$\ and $\delta y$ have units of 
radians. $\Delta$ is the geocentric distance. 

${\cal N}_i(x,y)$ is the number of CO molecules 
in the upper state of the transition sampled by the cells 
with Doppler velocities contributing to channel $i$: 
 
\begin{equation} 
 {\cal N}_i(x,y) =      \!\int_{x-\frac{\delta x}{2}}^{x+\frac{\delta x}{2}} 
                        \!\int_{y-\frac{\delta y}{2}}^{y+\frac{\delta y}{2}} 
                        \!\int_{-10\,L_{\rm CO}}^{\ 10\,L_{\rm CO}}\! n_{\rm CO}\,p_u\,{\rm H_i}\, dx\,dy\,dz, 
\end{equation} 
 
\noindent 
where $p_u$ is the relative population of the upper level of the transition, which depends on the radial distance to nucleus, and 
$n_{\rm CO}$ is from Eq.~\ref{haser_mere}. 
H$_i$, the function used to select velocities, is defined by: 
 
\begin{equation} 
 {\rm H}_i(x,y,z) = \left\{ 
\begin{array}{ll} 
1 & \textrm{if} \ v_z(x,y,z) \in [v_i-\frac{\delta v}{2}\, ; \, v_i+\frac{\delta v}{2}] \\ 
0 & \textrm{elsewhere} 
\end{array} 
\right. 
\end{equation} 
 
\noindent 
where $v_z(x,y,z)$ is the gas velocity projected onto the line of sight. 
The gas velocity is radial, and its amplitude is a Gaussian centred on 
$v_{\rm exp}$, which width is 
$2\sqrt{\ln(2)kT/m} $, to account for thermal broadening. $k$ is the Boltzman 
constant, $T$ the kinetic temperature and $m$ the CO molecular mass. 
 
Because the CO production rate is high in comet Hale-Bopp 
and a dense CO jet is present, optical depth effects 
need to be considered for the calculation of the CO $J$(2--1) brightness distribution (i.e., $F_i(x,y)$). 
They are not expected to affect significantly the ON--OFF spectra, but could 
be significant for the interferometric signals. The results presented in 
this paper were performed solving the full radiative transfer equation, 
as explained in \citet{boi+07}, assuming the local velocity dispersion 
to be thermal. 
 
A synthetic 47-channels ON--OFF spectrum is obtained by the 
convolution of $F_i$ with the antennas primary beam. 
 
For each channel $i$, the visibilities are defined by \citep[see e.g.,][ chap.~4,~\S4.1]{thompson}~: 
\begin{equation}\label{def-visibility} 
\mathcal{V}_i(\vec{\sigma}) = \frac{c}{\nu \delta v} \int_{4\pi} A(\vec{s}) F_i(\vec{s}) 
\exp(-\frac{2i\pi\nu}{c} \vec{\sigma} \cdot \vec{s}) d\Omega, 
\end{equation} 
where $\vec{\sigma}$ is the baseline vector for two antennas, with 
coordinates ($u$,$v$) in the $uv$-plane. \vec{s} is a vector in the 
sky plane which coordinates are $(x,y)$ in radian units. $A$ is the power pattern of the 
antennas, and $d\Omega$ is an element of solid angle on the sky. 
$\mathcal{V}_i(\vec{\sigma})$ is here in units of [W m$^{-2}$ Hz$^{-1}$] or 
janskys. 
Equation~\ref{def-visibility} can be approximated to~: 
\begin{eqnarray} 
\mathcal{V}_i(u,v) & = & \frac{c}{\nu \delta v} \int_{-\infty}^{+\infty} \int_{-\infty}^{+\infty} 
A(x,y) F_i(x,y) \nonumber \\ 
 & & \times \ \exp(-\frac{2i\pi\nu}{c}(ux+vy))\, dx dy. 
\end{eqnarray} 
Visibilities are computed with a Fast Fourier Transform (FFT) algorithm.

\begin{figure} 
\resizebox{\hsize}{!} 
{\includegraphics[angle=270]{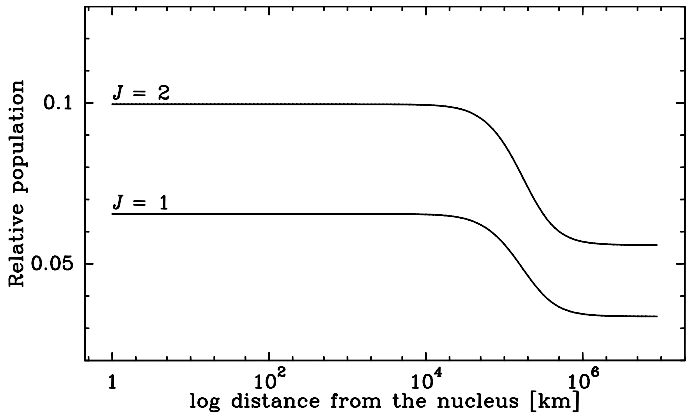}} 
\caption{Relative populations of the CO rotational levels $J=1$ and $J=2$, as function of distance from the nucleus. A kinetic temperature of 120 K is 
assumed.} 
\label{popu} 
\end{figure}

The population of the rotational levels ($p_u$) is derived from an 
excitation model which takes into account collisions with H$_2$O and 
IR radiative pumping of the $v$(1--0) CO vibrational band 
\citep{cro83,cro87}. This model provides populations $p_u$ as 
a function of radial distance $r$, given a H$_2$O density law with $r$. 
For simplicity, we assumed an isotropic H$_2$O coma, and $p_u$ only depending 
upon $r$.  The collisional CO--H$_2$O cross-section is taken 
equal to $\sigma_c = 2 \times 10^{-14}$ cm$^2$ \citep{biver99b}, and the 
H$_2$O production rate is 
$Q_{\rm H_2O} = 10^{31}$~s$^{-1}$ \citep{colom99}. In the simulations 
presented in Sect.~\ref{jets}, 
a kinetic temperature $T$ = 120 K is used (see Sect.~\ref{sect-onoff} 
for further discussion). The evolution of the population 
of the CO $J = 2 $ and $J = 1 $ levels is shown in Fig.~\ref{popu}. 
The beam size of $20.9\arcsec$ for CO $J$(2--1) corresponds to 
$r$ $\sim 10\,000$\,km in the coma. Most CO molecules within the 
field of view are in local thermal equilibrium. 
 
\begin{table} 
\caption{Model parameters. \label{param}} 
\begin{tabular}{lll} 
\hline 
\multicolumn{3}{c}{\bf fixed parameters} \\ 
\hline 
 heliocentric distance & 
$r_{h}$ & 
 0.989 AU\\ 
 geocentric distance & 
$\Delta$ & 
 1.368 AU\\ 
 gas expansion velocity$^{\mathrm{(a)}}$ & 
${v}_{\rm exp}$ & 
 1.05\,km\,s$^{-1}$ \\ 
 total CO production rate$^{\mathrm{(a)}}$ & 
$Q_{\rm CO}$ & 
 2.0\,$\times$\,10$^{30}$~s$^{-1}$ \\ 
 CO photodissociation rate$^{\mathrm{(b)}}$  & 
$\beta_{\rm CO}$ & 
 $7.50\,\times\,10^{-7}\,r_h^{-2}$ \\ 
 nucleus rotation period & 
$P$ & 
11.35~h \\ 
\hline 
\multicolumn{3}{c}{\bf free parameters} \\ 
\hline 
\multicolumn{3}{l}{ 
\begin{tabular}{ll} 
fraction of CO in the coma outgassed in the jet & \fco \\ 
jet latitude & $\ell$  \\ 
jet half-width & $\Psi$ \\ 
 rotation axis aspect angle & \thw \\ 
 rotation axis position angle & \paw \\ 
\end{tabular} 
}\\ 
\hline 
\end{tabular} 
\begin{list}{}{} 
\item[$^{\mathrm{(a)}}$] \citet{biver99} 
\item[$^{\mathrm{(b)}}$] \citet{huebner92} 
\end{list} 
\end{table}  

\begin{figure} 
\resizebox{\hsize}{!} 
{\includegraphics{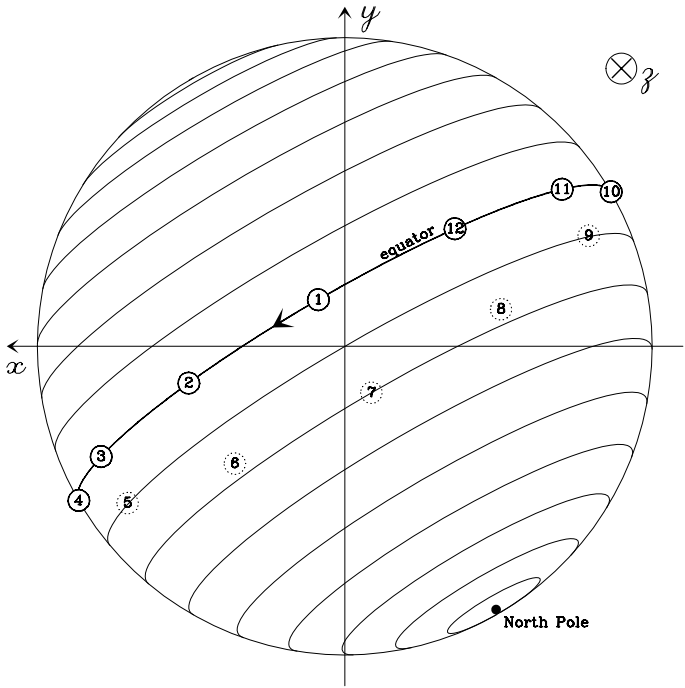}} 
\caption{Schematic view of comet Hale-Bopp nucleus on 11 March, 1997, 
as seen from the Earth assuming 
\paw=210\degre\ and \thw=80\degre for the spin orientation. East is on the 
left. 
The latitudes are shown by steps of 10\degre\ and the arrow shows the rotation 
direction. 
The symbols represent the series of jet positions used to make a composite 
$uv$-table for an equatorial jet. 
Plain (respectively dotted) symbols mean that the jet is on the visible 
(respectively hidden) side of the nucleus. 
\label{rot}} 
\end{figure}

Given the evolving coma, a full simulation of the observations 
would require to compute for each one minute scan the visibilities 
corresponding to the current state of the coma and to their 
$uv$-coverage. In order to limit the computer time, we modelled a 
whole nucleus revolution ($P$ = 11.35 h) by 
12 snapshots (see Fig.~\ref{rot}). We tested the validity of the 
approach by verifing that calculations with an increased time sampling 
(namely 36 snapshots) provide similar results. Between 2 snapshots $i$ and 
$i+1$, the jet direction changed following 
$(\theta_0,\phi_0)_{i+1}= \vec{R}_{\omega}(\theta_0,\phi_0)_i$, 
where $\vec{R}_{\omega}$ is the rotation matrix for a span of 
$P/12$. Providing the initial jet longitude at time $t$ 
corresponding to snapshot $i$=1 is fixed, a composite $uv$-table 
can be computed with the $uv$-coverage of the observations. The 
numerical code computes twelve composite $uv$-tables, each of them 
corresponding to different initial jet longitudes spaced by 
360\degre$/12$. The longitude origin is chosen so that the 
sub-terrestrial point on the nucleus surface is at a longitude of 
0\degre. For illustration, Fig.~\ref{rot} shows the twelve jet 
positions for a jet at a longitude of 0\degre at the time of the 
first snapshot. The model also computes synthetic ON--OFF spectra 
for each snapshot.

\section{Jet morphology analysis\label{jets}} 
 
In this Section, the ON--OFF and interferometric velocity shift 
curves observed for CO $J$(2--1) (Figs~\ref{evol-velo} 
and~\ref{evol-velo-int-mod}), and the time evolution of the 
visibilities (Fig.~\ref{visi-obs}) will be analysed with the model 
presented in the previous Section to constrain its free 
parameters. Because of limited signal-to-noise ratio, it was not 
possible to make this analysis for $J$(1--0) observations. For the 
spin axis orientation defined by its aspect angle \thw and 
position angle \paw, we restricted our study to the mean values 
found in the literature (see Table~\ref{polpos}): \thw~=~60 to 
90\degre, and \paw~=~200 to 230\degre. The jet width $\Psi$ was 
tested from 1\degre\ to 90\degre, the jet latitude $\ell$ from 
0\degre\ to 90\degre\ North, and \fco from 10\% to 80\%. 
 
\subsection{ON--OFF velocity shift curve} 
\label{on-off-vsc} 
 
\begin{figure} 
\resizebox{\hsize}{!} 
{\includegraphics[angle=270]{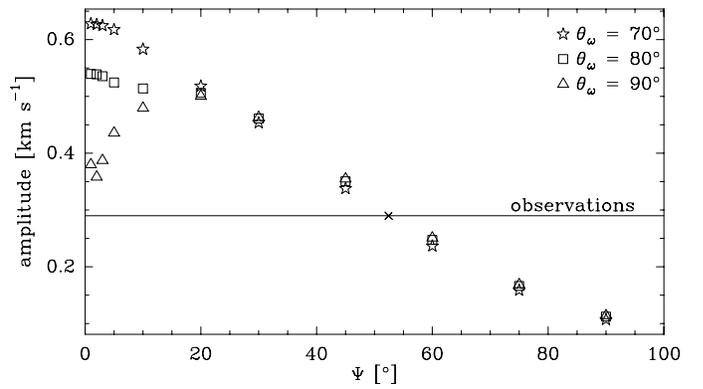}} 
\caption{Evolution of the amplitude of the velocity shift curve \amp with \thw and $\Psi$, for \fco=~66\% and \lat = 0\degre. 
The cross indicates the value of $\Psi$ required to fit the observations 
($\mathcal{A}$ = $0.29 \pm 0.03$ ~km\,s$^{-1}$) for 
\thw~=~70--90\degre and \fco~=~66\%. 
\label{evolpsi}} 
\end{figure} 
 
As discussed in Sect.~\ref{sect-onoff}, data shown in Fig.~\ref{evol-velo} 
are well fitted by a sinusoid with a period 
 $P=11.35$\,h, an amplitude $\mathcal{A}$ = 0.29~km\,s$^{-1}$, and 
centred on $v_0=-0.05$~km\,s$^{-1}$. We can also define a phase $t_0 = 
11.75\pm 0.12$~UT, that corresponds to the time when the velocity shift 
is equal to  $v_0$ on the increasing side of the curve. 
We used these three parameters ($\mathcal{A}$, $v_0$, $t_0$) as 
criterions for selecting the models that could explain the observations. 
Note that the position angle \paw of the spin axis 
has no influence on the velocity shift. 
 
The phase $t_0$ is only ruled by the initial longitude of the jet. From $t_0$ derived 
from the observations, the best initial longitude is 300\degre at 3~h~47~UT. 
 
\begin{figure} 
\resizebox{\hsize}{!} 
{\includegraphics[angle=270]{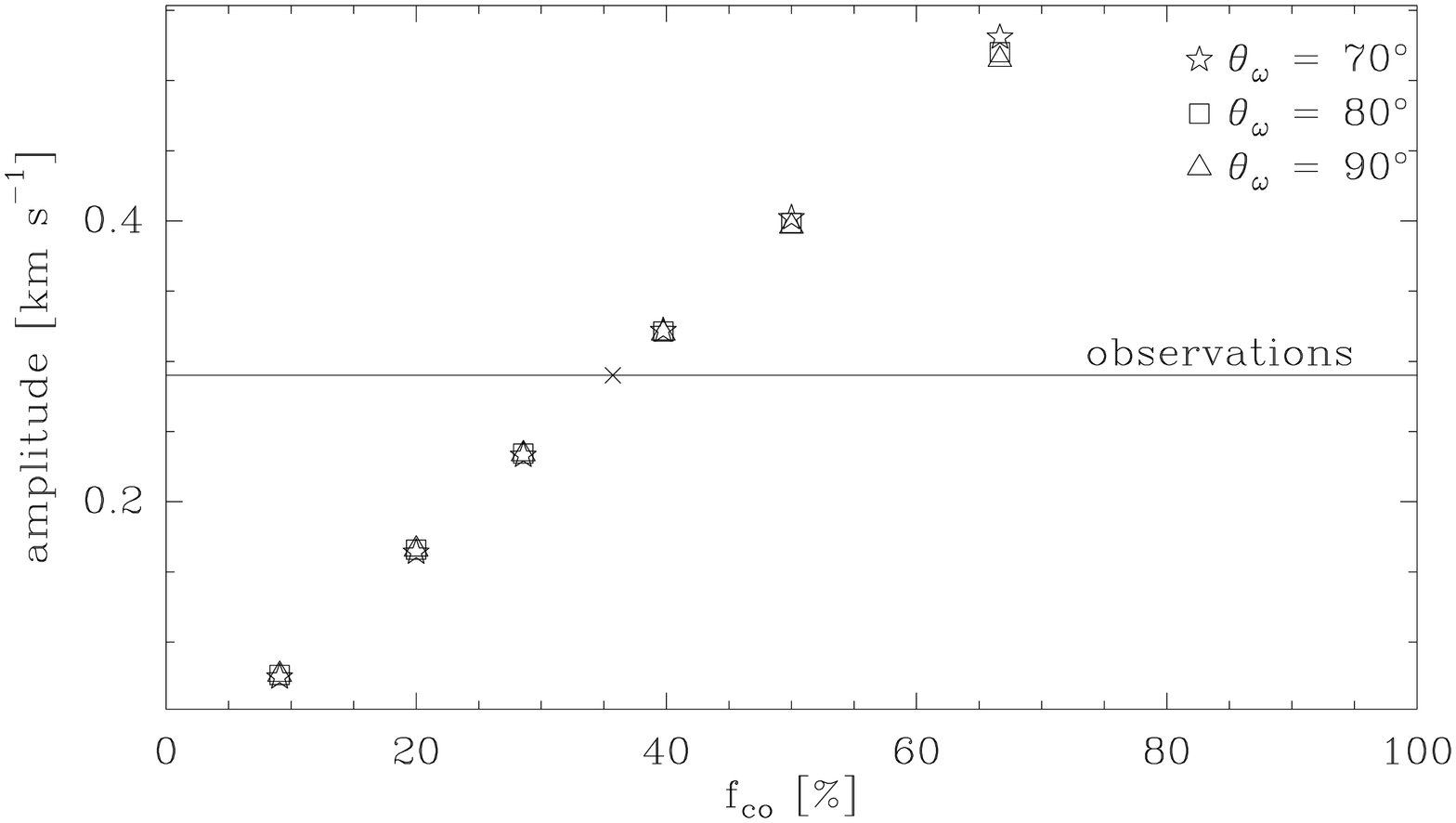}} 
\caption{Evolution of the amplitude of the velocity shift curve \amp with 
\thw and \fco, for $\Psi$\,=\,20\degre and \lat = 0\degre. 
The cross indicates the value of \fco required to fit the observations ($\mathcal{A}$ = 0.29 $\pm$ 0.03~km\,s$^{-1}$) for 
\thw~=~70--90\degre and $\Psi$\,=\,20\degre. 
\label{evolfco}} 
\end{figure}

\begin{figure} 
\resizebox{\hsize}{!}{%
\includegraphics[angle=270]{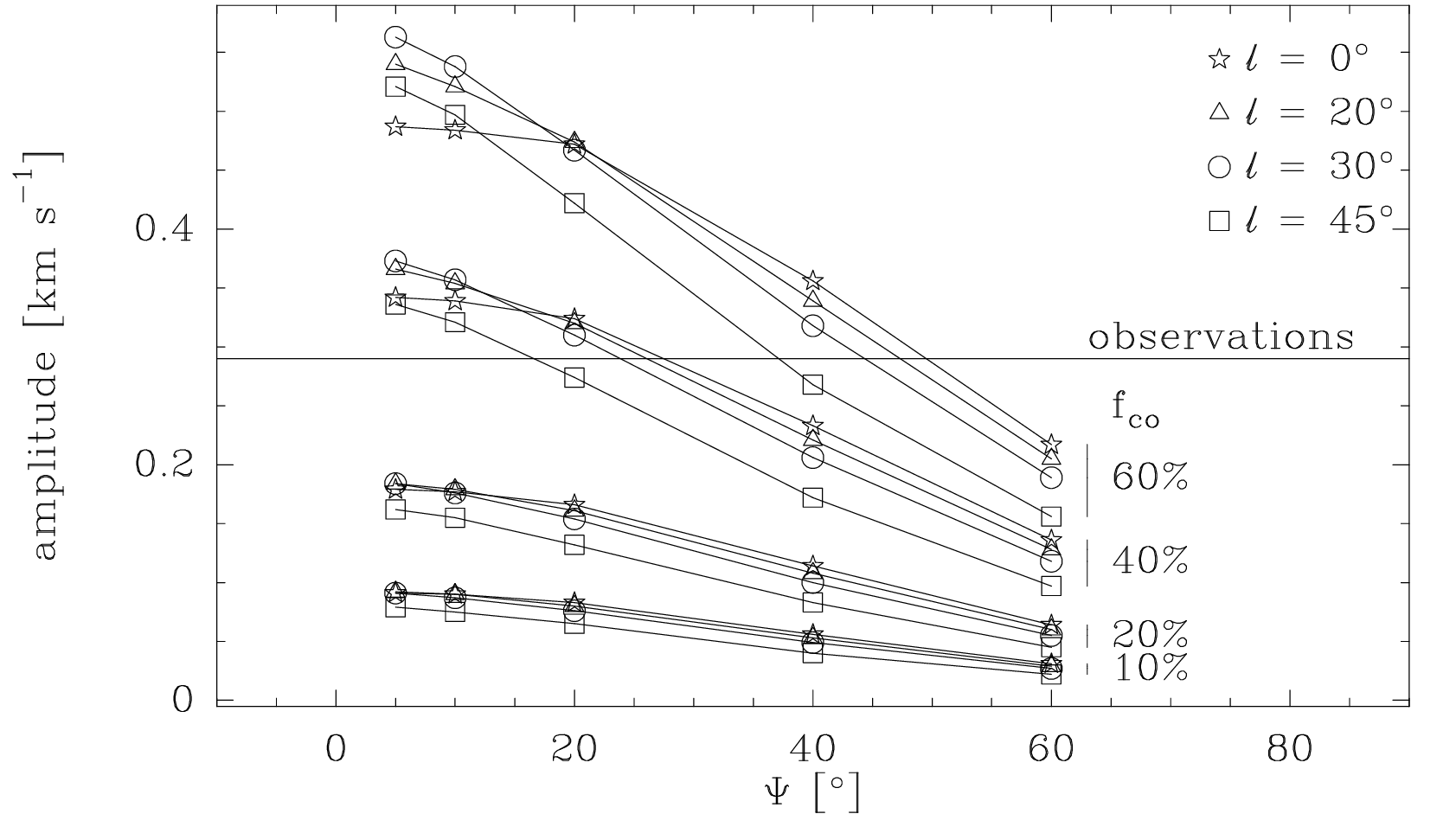}} 
\vspace{+0.2cm} 
\resizebox{\hsize}{!}{%
\includegraphics[angle=270]{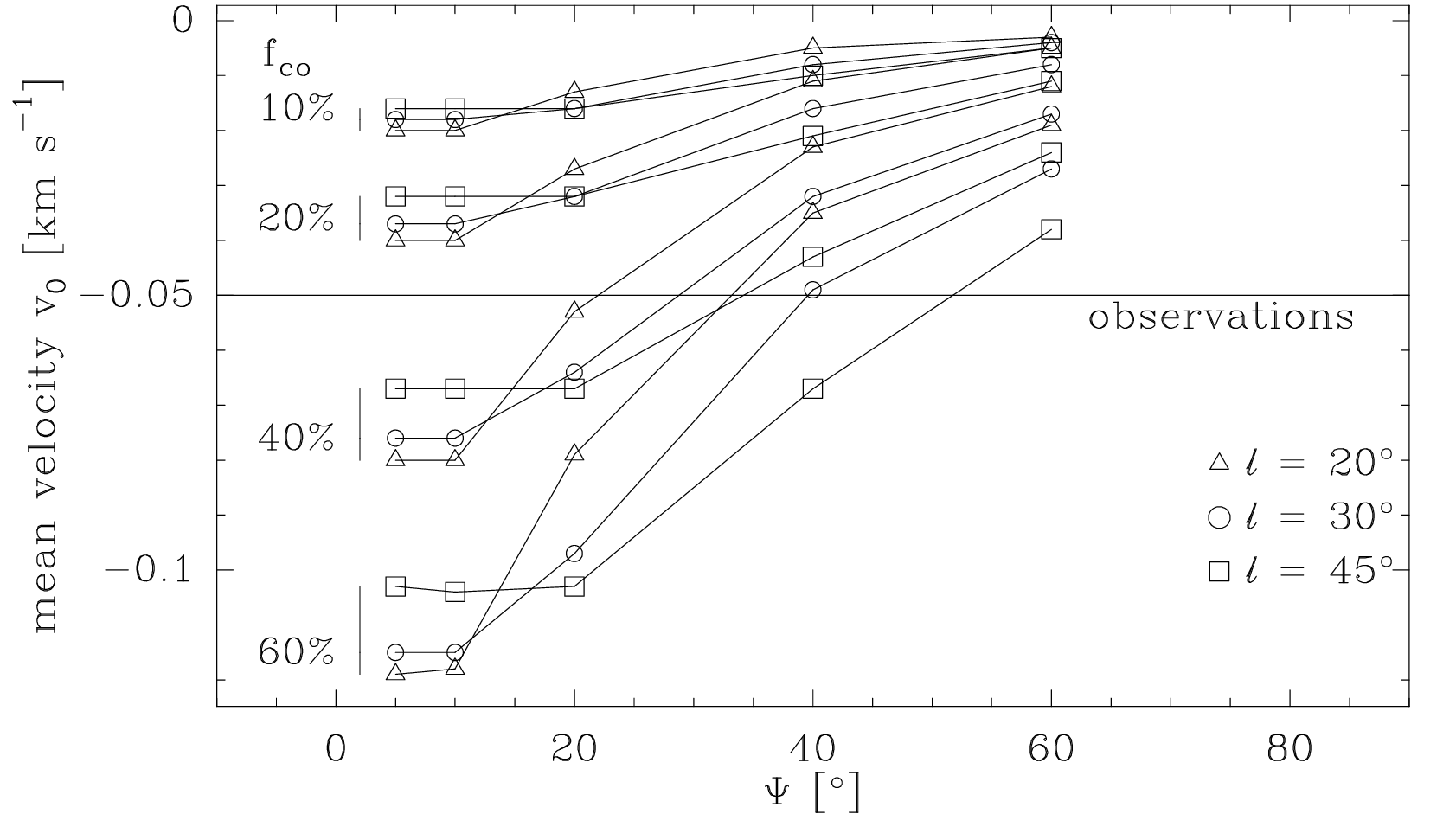}} 
\caption{Evolution of \amp (top) and  $v_0$ (bottom) with $\Psi$ for several \fco and \lat. 
Simulations have been done 
with \thw~=~80\degre and \paw~=~210\degre.Observed values are 
$\mathcal{A}$ = 0.29 $\pm$ 0.03~km\,s$^{-1}$ and $v_0$ = -0.05 $\pm$ 0.01~km\,s$^{-1} $ ) 
\label{evolamp-fcopsilat}} 
\end{figure}

At first order, \amp is governed by \fco and $\Psi$ : it increases when \fco 
increases or $\Psi$ decreases.  This behaviour is easily explained. If \fco 
increases -- all other parameters remaining unchanged -- there is more signal 
coming from the jet falling into the same number of spectral channels.  Then, 
the velocity shift increases, so does the amplitude of the velocity shift curve. 
In a similar way, when $\Psi$ 
decreases -- keeping \fco constant -- an equal amount of signal coming from the jet 
is falling into a larger number of spectral channels. This results in reducing the 
velocity shift and also \amp.  This is illustrated in Figs~\ref{evolpsi} 
and~\ref{evolfco}, 
which show the evolution of \amp with \fco and $\Psi$.  Moreover, \amp is not 
sensitive to the aspect angle \thw, except for small $\Psi$'s, as observed in Figs~\ref{evolpsi} and 
\ref{evolfco}.  Furthermore, we note that the jet latitude $\ell$ has little 
influence upon the amplitude, within the limits where we tested it ($|\ell|<45$\degre) (Fig.~\ref{evolamp-fcopsilat}). 
To conclude, the observed amplitude of 0.29\,km\,s$^{-1}$ can be fitted by many 
(\fco,$\Psi$,$\ell$) combinations. 
 
The mean velocity $v_0$ of the simulated curves depends mainly on the jet latitude 
$\ell$.  An equatorial jet always produces a curve centred on 0\,km\,s$^{-1}$, whatever 
\thw. For \thw~$<$~90\degre, a 
jet with a northern (resp. southern) latitude produces a curve centred on a 
negative (resp. positive) velocity. Again, this behaviour is understood.  With 
\thw~$<$~90\degre, the North pole is pointing towards the Earth (see Fig.~\ref{rot}). 
As a result, a northern jet is more often directed towards the Earth than a 
southern one.  Opposite effects are obtained for \thw~$>$~90\degre, while 
\thw~=~90\degre\ (rotation axis in the plane of the sky) always produces a curve 
centred on 0\,km\,s$^{-1}$, irrespective of the sign of the latitude. 
The more the rotation axis is far from the plane of 
the sky, the more the velocity shift curve is shifted.  Furthermore, 
the velocity shift curve is all the more shifted as $\Psi$ is small and \fco 
is large (Fig.~\ref{evolamp-fcopsilat}). 
This study leads us to the conclusion that, here again, many combinations (\fco,$\Psi$,$\ell$) 
are able to reproduce the observed $v_0$. However, it shows that only a 
northern jet will be able to fit the observations. 

\begin{figure*} 
\centering 
\includegraphics[angle=270,width=17cm]{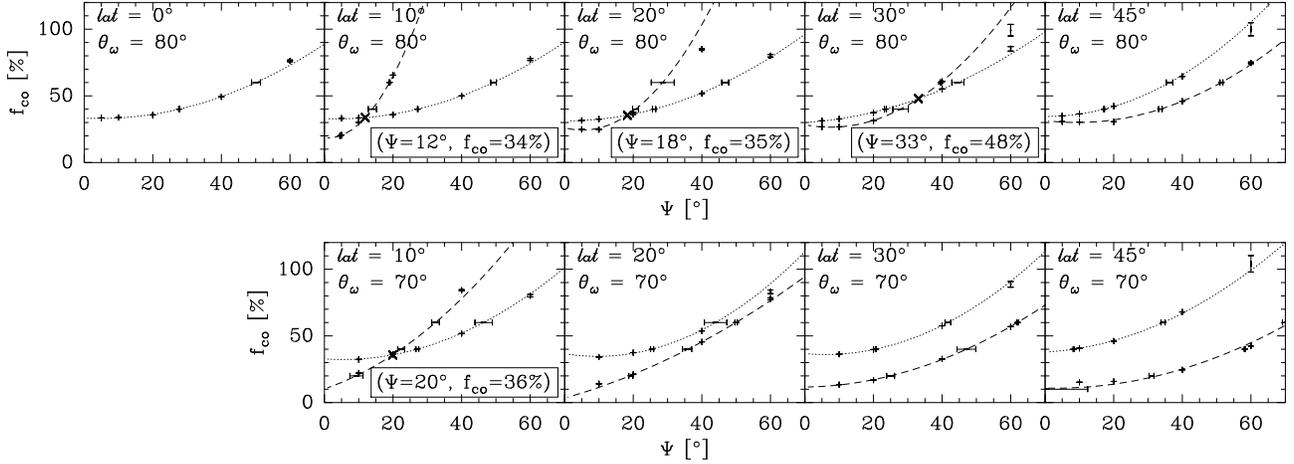} 
\caption{Couples (\fco,$\Psi$) that reproduce the 
amplitude \amp~=~0.29 $\pm$ 0.03~\kms (dotted curves) and the mean velocity 
$v_0$~=~-0.05 $\pm$ 0.01~\kms 
(dashed curves) for several couples of (\thw,\lat). The intersections of 
the curves give, for each (\thw,\lat) combination, the only couple 
(\fco,$\Psi$) that reproduces \amp and $v_0$. 
\label{fcopsi}} 
\end{figure*}  
 
For a given (\thw,$\ell$,\fco) parameter set, it is possible to 
find the jet width $\Psi$ that is able to reproduce the amplitude  $\mathcal{A} 
= 0.29$~km\,s$^{-1}$. Figure \ref{fcopsi} shows in dotted curves the locus of the 
couples (\fco,$\Psi$) that reproduce the right amplitude $\mathcal{A}$ for several 
fixed (\thw,$\ell$) values.  The same method is employed to determine the couples 
(\fco,$\Psi$) that reproduce the right velocity centre $v_0 = -0.05$~\kms (dashed curves in 
Fig. \ref{fcopsi}).  For each set (\thw,$\ell$), the 
intersection of the dotted and the dashed curves gives the only couple (\fco,$\Psi$) 
that reproduces $\mathcal{A}$ and $v_0$. We made these computations for latitudes 
between 0\degre and 45\degre, and for \thw~=~70\degre and 80\degre. Computations 
for \thw~=~90\degre were useless because they provide $v_0=0$~\kms. 
 
The combinations (\thw,\fco,$\Psi$,$\ell$) selected by this study are summarized 
in Table~\ref{selected-couples}.  We note that the jet strength \fco is typically between 
35 and 50\%.  The jet is located on the northern hemisphere at a latitude 
between 0\degre and 45\degre (both excluded) for 
\thw~=~80\degre, and between 0\degre and 20\degre (both excluded) for \thw~=~70\degre. 
 
\begin{figure} 
\begin{center} 
\includegraphics[bb = 18 17 555 272,angle=270, width = 6cm, clip=true]{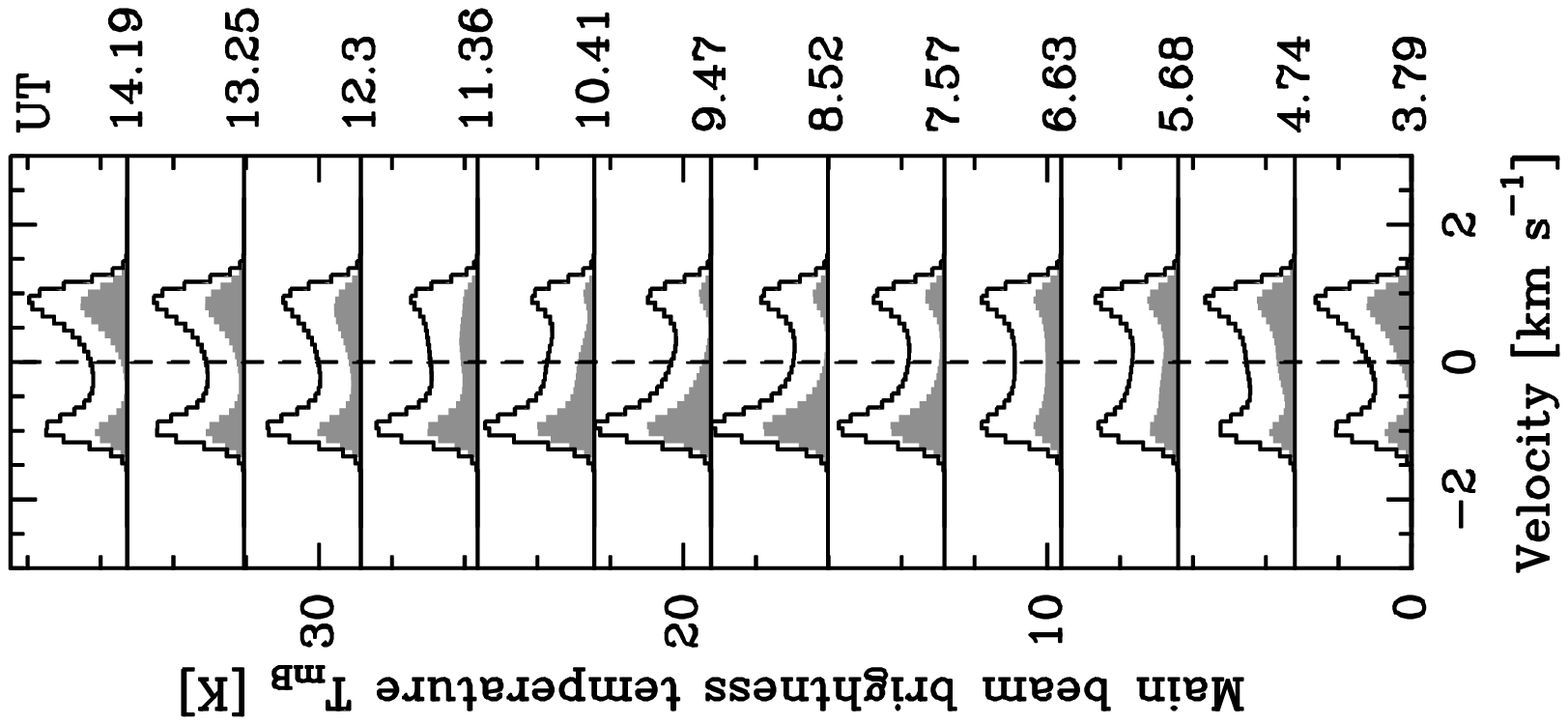} 
\end{center} 
\caption{Synthetic CO 230 GHz ON--OFF line profile as a function 
of UT time on 11 March for parameter set (3) of 
Table~\ref{selected-couples} (\thw~=~80\degre, \paw~=~210\degre, 
$\Psi$~=~18.3\degre, \lat~=~20\degre and \fco~=~35.5\%). The total 
spectrum (jet+isotropic contribution) is shown by thick lines. The 
grey spectrum shows the contribution of the CO jet.} 
\label{sp-mod} 
\end{figure}

The velocity shift curve obtained for parameter set (3) is shown 
in Fig.~\ref{evol-velo-int-mod}, together with the observed curve. 
We note the almost perfect match between model and observations. 
The corresponding synthetic line profiles are shown in 
Fig.~\ref{sp-mod}. At any time, the jet contributes to both blue 
and red channels, with varying relative contributions, due to its 
spiral shape.

\begin{table} 
\caption{Selected sets of parameters \thw, \lat, $\Psi$ and \fco 
reproducing the velocity shift curve of the ON--OFF observations. 
\label{selected-couples}} \centering 
\begin{tabular}{ccccc} 
\hline 
set &\thw & \lat & $\Psi$ & \fco  \\ 
\hline 
(1) &70\degre & 10\degre & 19.9\degre & 35.7\% \\ 
(2) &80\degre & 10\degre & 11.9\degre & 33.6\% \\ 
(3) &80\degre & 20\degre & 18.3\degre & 35.5\% \\ 
(4) &80\degre & 30\degre & 33.0\degre & 47.8\% \\ 
\hline 
\end{tabular} 
\end{table}

\subsection{Interferometric velocity shift curves}

In this Section, we study the velocity shift curves observed for 
the individual baselines. Figure~\ref{evol-velo-int-mod} shows 
model results with the parameter set (3) (\thw~=~80\degre, 
\paw~=~210\degre, $\Psi$~=~18.3\degre, \lat~=~20\degre and 
\fco~=~35.5\%; Table~\ref{selected-couples}). Other parameter sets 
of Table~\ref{selected-couples} give similar curves. Modelled 
curves are periodic functions, with a period equal to $P$. They 
mimic sinusoidal curves, though significant deviations from a 
sinusoid are observed. This is because line shifts measured on 
$\mathcal{V}(u,v)$ spectra are $(u,v)$ dependent: stronger jet 
contrast appears in specific regions due to spatial filtering. 
Then, due to the combination of Earth and jet rotation, regions 
with more or less jet contrast are sampled by the individual 
baselines. Simulations show that these curves evolve toward a true 
sinusoid when the jet width $\Psi$ is increasing (\fco kept 
constant), due to smaller jet contrast. These curves change when 
varying the jet parameters in the same way than does the ON--OFF 
curve. Changing the spin axis parameters by $\pm$ 10\degre does 
not affect much the curves. 
 
\begin{figure} 
\resizebox{\hsize}{!} {\includegraphics[bb=60 43 536 690,clip=true]{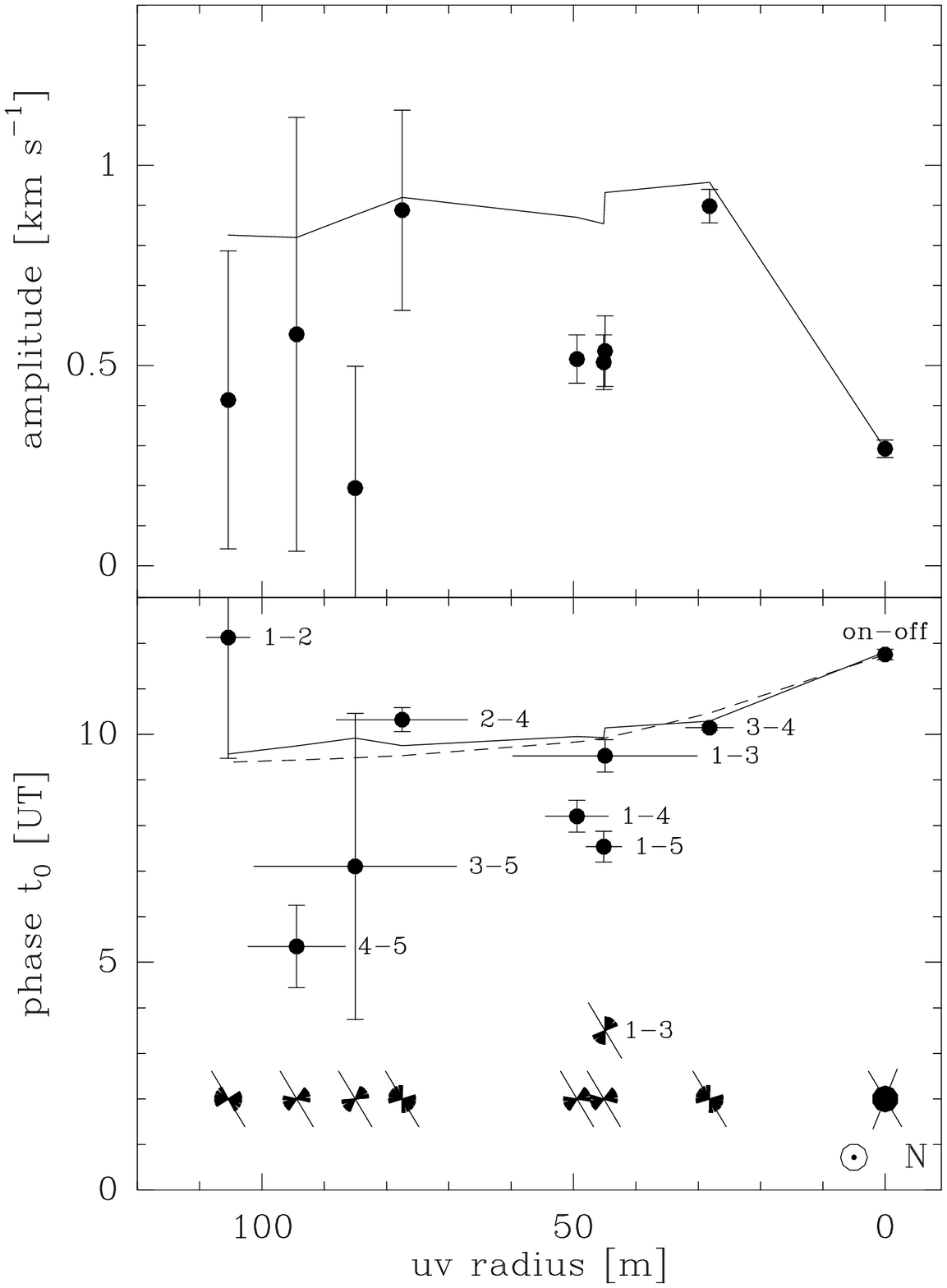}} 
\caption{Phase $t_0$ 
(bottom) and amplitude \amp (top) of the velocity shift curves as 
a function of $uv$-radius. Dots with error bars are the data. The 
range of $uv$ radius spanned by the baselines is shown with 
horizontal lines in the bottom figure. The plain curve shows model 
calculations with \thw~=~80\degre, \paw~=~210\degre, 
$\Psi$~=~18.3\degre, \lat~=~20\degre and \fco~=~35.5\%. The dotted 
curve in the bottom figure assumes that the elapsed time between 
jet detection by baseline at $uv$-radius $\sigma_1$ and jet 
detection by baseline at $\sigma_2$ (or in ON--OFF spectrum at 
$\sigma_2$ = 15 m) is equal to ($\sigma_2$-$\sigma_1$)/2$v_{\rm 
exp}$, with $v_{exp}$ = 1.05 km s$^{-1}$. The reference time is 
$t_0$ measured on the ON--OFF velocity shift curve. The 
orientation of the baselines fringes during the course of the 
observations is shown at the bottom, together with the spin axis 
direction in the (RA, Dec) plane: e.g., the field of view of 
baseline 3--4 has its long dimension (reflecting the primary beam 
of the antenna) along the spin vector of the comet, and its short 
dimension (related to the baseline length) perpendicular to it. 
The Sun direction in the (RA, Dec) plane is also indicated.} 
\label{velocityshift} 
\end{figure} 
 
\begin{figure} 

\resizebox{\hsize}{!} {\includegraphics[angle=270,width=8cm,clip=true]{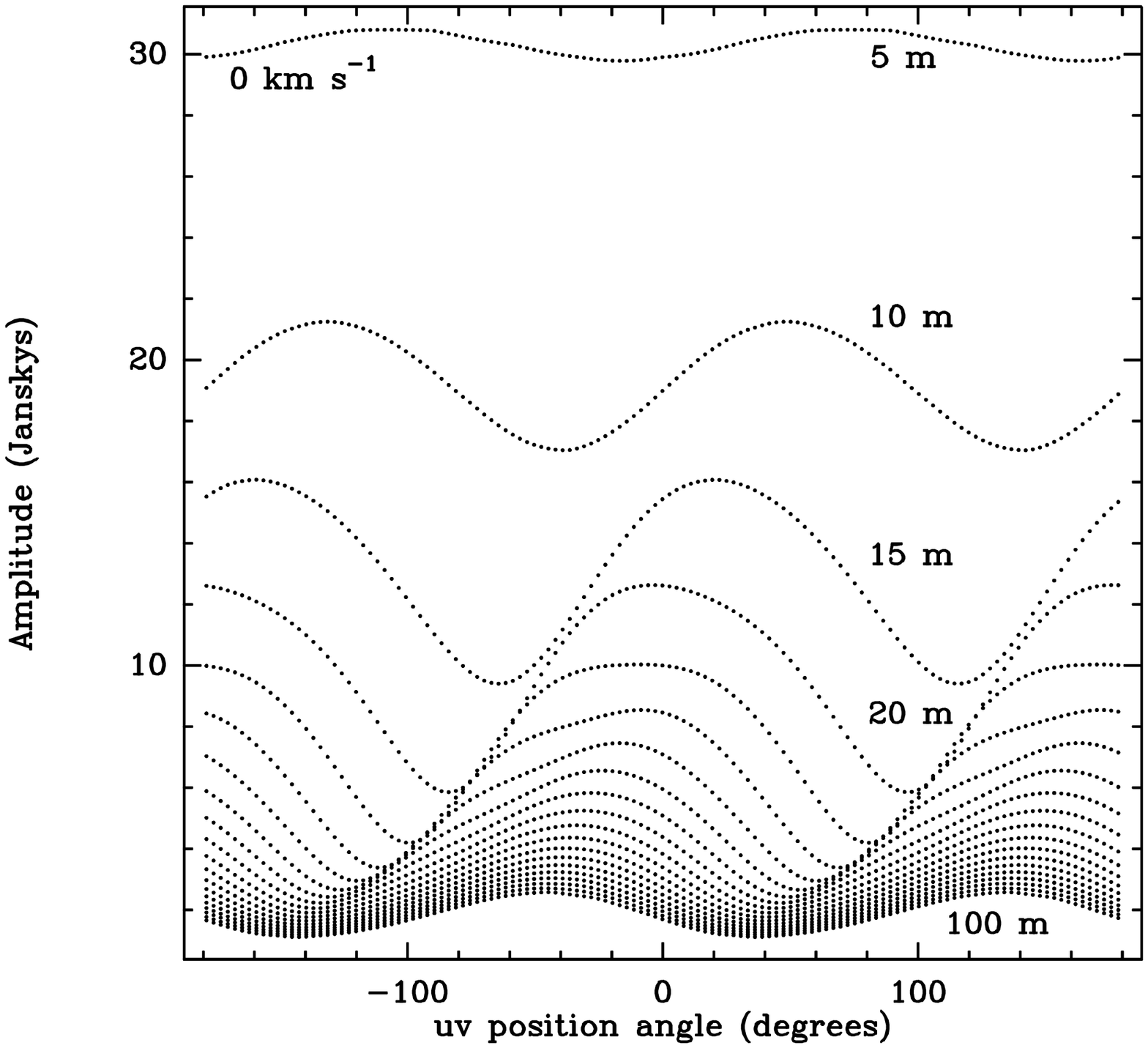}} 
\caption{Visibilities of the CO 230 GHz line (central chanel) as a 
function of baseline orientation in the $uv$ 
plane and baseline length ($\sigma$ from 5 to 100 m by step of 5 m) 
at a given time. The rotation axis 
($P$ = 11.35 h) is along the line of sight (\thw~=~0\degre). The jet is 
equatorial, has an aperture $\Psi$~=~30\degre, and 
its direction at the nucleus surface is at a position angle of 60\degre 
in the plane of the sky (i.e., $\theta_0$~=~30\degre). Other parameters  
are $Q_{\rm CO}$ = 1 $\times$ 10$^{30}$ s$^{-1}$, $f_{\rm CO}$ = 1, with 
assumed optically thin conditions.} 
\label{visi-po} 
\end{figure}

The modelled velocity shift curves for the different baselines are 
not in phase, with $t_0$ (defining the phase, see 
Sect~\ref{on-off-vsc}) increasing with decreasing baseline length 
(Fig.~\ref{velocityshift}). We expect a phase offset due to  
the spiral shape of the jet. Indeed, with respect to long baselines, 
short baselines probe molecules in more distant regions of 
the spiral. Hence, they sample molecules released in average at 
earlier times. In addition,  baselines of different 
length (even if they are parallel) record the maximum signal from the 
jet at different times
due to the curvature of the jet. This can be understood from   
Fig.~\ref{visi-po}, which plots the amplitude of the visibility 
as a function of the orientation and length of the baselines for
a simple geometry (rotation axis along the line of sight and  
equatorial jet) and at a given time. The combination of both effects 
introduces a phase offset in the velocity shift curves. 
The delay between two baselines in the 
velocity shift curves represents the elapsed time between the jet 
detection by one baseline and its detection by the following one. 
Note that, given the large curvature of the 
spiral (molecules travel $r$ $\sim$ 10$^4$ km when the nucleus 
rotates by 90\degre), only its innermost part contributes 
significantly to the detected signal. The dashed curve in 
Fig.~\ref{velocityshift} shows the evolution of $t_0$ with 
$uv$-radius, assuming that $t_0$ varies linearly with 
$\sigma$/$v_{exp}$: it follows approximately the $t_0$ curve 
computed by the model (plain curve).      

The comparison between modelled and observed phases $t_0$ and 
amplitudes \amp shows that there is relatively good agreement for 
some baselines and strong discrepancies for others 
(Fig.~\ref{velocityshift}). Good agreement for both  $t_0$ and 
\amp is obtained for baselines 3--4 and 2--4. Phase $t_0$ is well 
reproduced for baseline 1--3 (but not the amplitude). Strong 
discrepancy (by 3 h for $t_0$ and a factor of almost 2 in \amp) is 
observed for baselines 1--4, 1--5 and possibly 4--5 (errorbars are 
large for this baseline). It is striking to note that good 
agreement is obtained for baselines with their field of view 
aligned along the spin vector, while discrepancies are observed 
for baselines with their field of view perpendicular to the spin 
vector.  Clearly our model is too simple to reproduce all 
observational characteristics. We did not find any simple 
explanation for these discrepancies. Velocity acceleration inside 
the jet changes $t_0$ evolution with baseline in the opposite way: 
it would result in a flatter increase with decreasing baseline 
length than obtained with a constant velocity. The presence of 
other CO evolving structures in Hale-Bopp coma is the likely 
explanation (Sect.~\ref{otherstructure}).

\subsection{Visibilities} 
\label{visi} 
 
Figure~\ref{visi-obs} shows the time evolution of the 
visibilities $\bar{\mathcal{V}}$ plotted as a function of the $uv$-radius 
$\sigma$ (as defined in Sect. 2.3, $\bar{\mathcal{V}}$ refers to 
the amplitude of the visibilities integrated over velocity). A 
least-squares fit to these data gives $\bar{\mathcal{V}}(\sigma) 
\propto \sigma^{-1.18\pm0.02}$, to be compared to the 
$\sigma^{-1}$ variation expected for a parent molecule 
distribution and an optically thin line \citep{Boc09}. 
This trend can be explained by optical depth effects being more important for 
long baselines probing the inner coma. 
 
\begin{figure} 
\resizebox{\hsize}{!} 
{\includegraphics[angle=270]{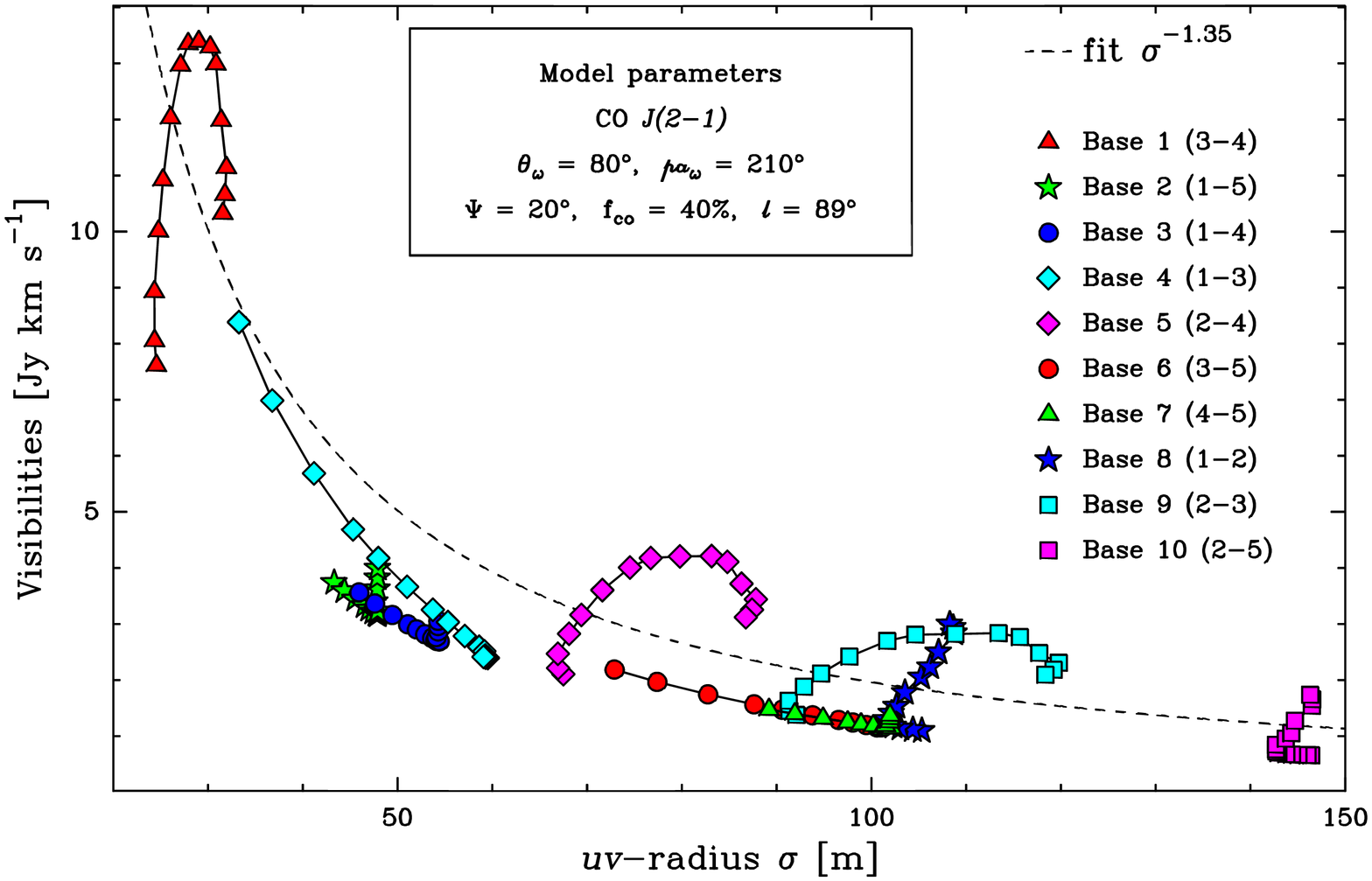}}
\caption{CO 230 GHz modelled visibilities for a high-latitude jet. 
 } 
\label{high-lat} 
\end{figure}

Modulations are observed around this fit: they trace variations 
of the brightness distribution sampled by the 
individual baselines as the baselines and jet are rotating. These 
modulations are characterized by their shape and their amplitude. The shape depends 
essentially upon the rotation axis position angle \paw and 
the jet latitude $\ell$. For example, a high-latitude jet would 
result in strong modulations for baselines 3--4, 2--4 and 2--3, and 
no modulations for baselines 1--3 and 
1--4  which scan regions along the equator (Fig.~\ref{high-lat}). 
The reverse is expected for 
a high-latitude jet and \paw at 90\degre from the nominal \paw of Hale-Bopp 
rotation axis. Qualitatively, the observed modulations exclude a high-latitude jet, as well as they 
exclude a rotation axis position angle much different than the \paw 
derived from visible observations. This confirms the conclusion obtained from 
the time evolution of the photometric centres (Sect.~\ref{inter-data}), sensitive 
to both the amplitude and phase of the visibilities. 
 
\begin{figure} 
\resizebox{\hsize}{!} 
{\includegraphics[angle=270]{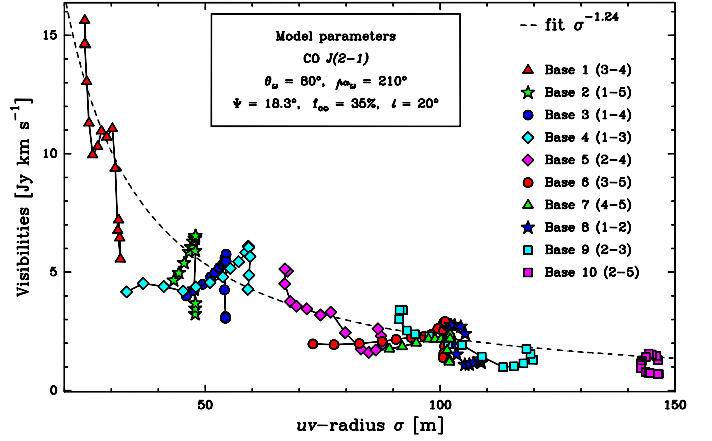}} 
\caption{CO 230 GHz modelled visibilities for the jet and rotation axis 
parameters which reproduce at best the observations (set (3) with \paw 
= 210\degre).} 
\label{visi-model} 
\end{figure} 

\begin{figure} 
\resizebox{\hsize}{!} 
{\includegraphics[angle=270]{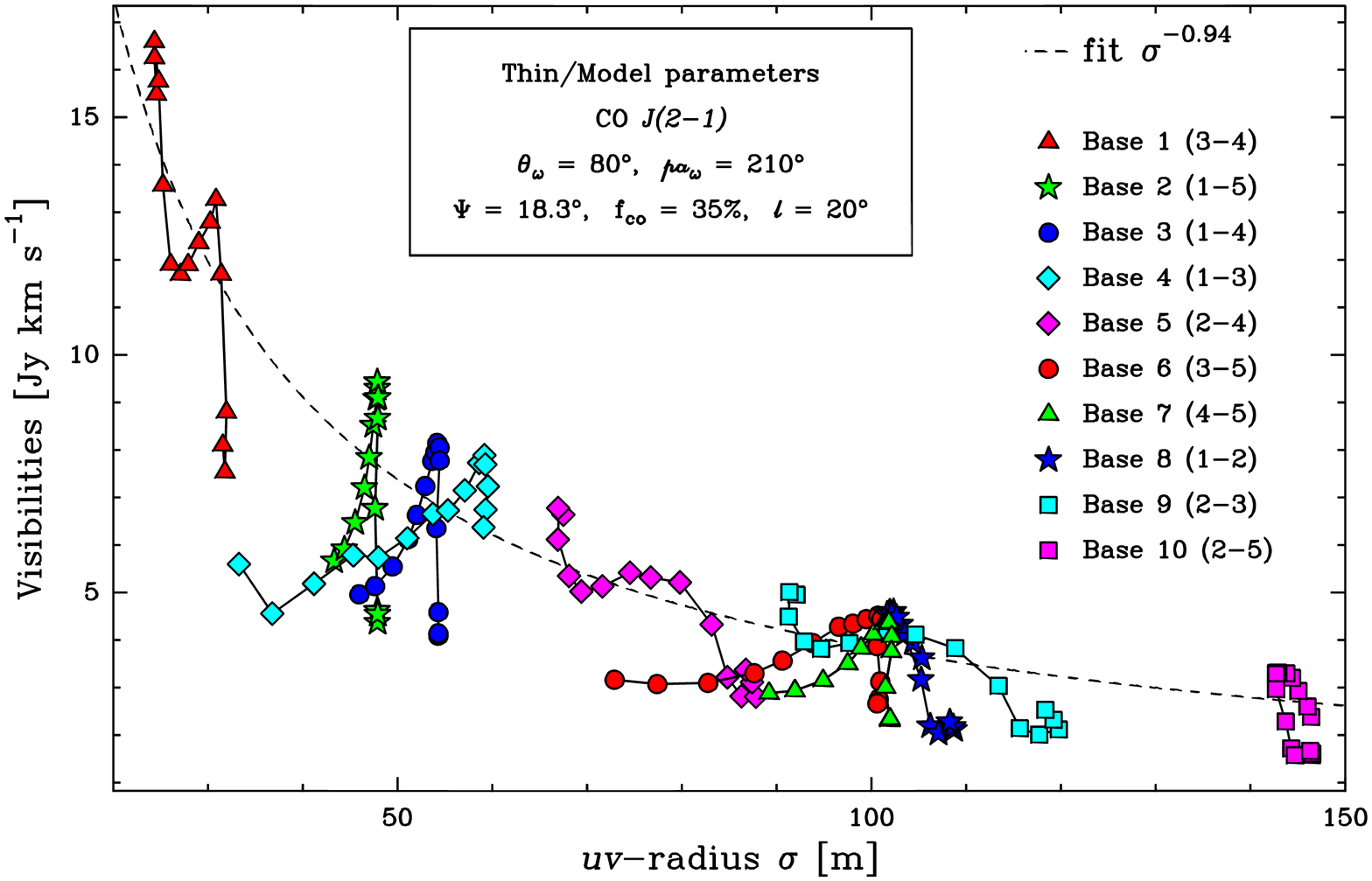}}

\caption{CO 230 GHz modelled visibilities for set (3) with \paw 
= 210\degre assuming an optically thin line.} 
\label{visi-model-alter} 
\end{figure}

We compared the observed visibilities to those computed with the 
model with jet parameters (2), (3) and (4) given in 
Table~\ref{selected-couples} (those with \thw~=~80\degre and 
\paw~=~210\degre). The lowest $\chi^2$ (reduced $\chi^2_{N-n}$ = 5.1  
for $N$ = 150 data points and $n$ = 5 free parameters)  
was obtained for the set of 
parameters (3) with $\Psi$~=~18.3\degre, \lat~=~20\degre and 
\fco~=~35.5\%. Simulations with parameters (2) (respectively (4)) 
show larger (resp. lower) modulations than observed, and $\chi^2$ 
values 60\% (resp. 22\%) larger than with parameters (3). Using 
jet parameters (3), we also made simulations with \thw~=~60 and 
70\degre and \paw~=~200, 220 and 230\degre. The  $\chi^2$ was 
minimized for \paw~=~220\degre ($\chi^2_{N-n}$ = 3.8), while \thw has no 
significant influence on the visibilities. However, \paw~=~210\degre explains 
better the visibilities of the 3--4 baseline. Figure
\ref{visi-model} shows the modelled visibilities 
with parameters (3) and \paw~=~210\degre. Looking to the shape 
of the modulations, there is an overall agreement between model and 
observations though, admittedly, the agreement is not perfect. A 
plot of observed versus modelled visibilities shows that the best 
agreement is for baselines 3--4, 1--5, 1--4, 1--3 and 4--5. The 
largest discrepancies are for baselines 2--4, 1--2, and 2--5. For 
most baselines, jet detection (traced by amplitude increase) 
occurs $\sim$ 1 h earlier in the simulation than in the 
observations. In contrast, in the velocity shift curves, the 
simulated jet is late with respect to the observed jet for most 
baselines. This again shows that our model is too simple to 
explain satisfactorily the data.

The visibilities obtained when optical depth 
effects are neglected show the same temporal behaviour 
(Fig.~\ref{visi-model-alter}). However, as expected, modulations are 
more contrasted in the optically thin case.

The visibilities obtained with jet parameters (3) vary according to 
$\bar{\mathcal{V}}(\sigma) \propto \sigma^{-1.24}$, which is 
consistent, in first order, with the observed variation 
($\propto \sigma^{-1.18 \pm 0.02}$).

\begin{figure} 
\resizebox{\hsize}{!} 
{\includegraphics{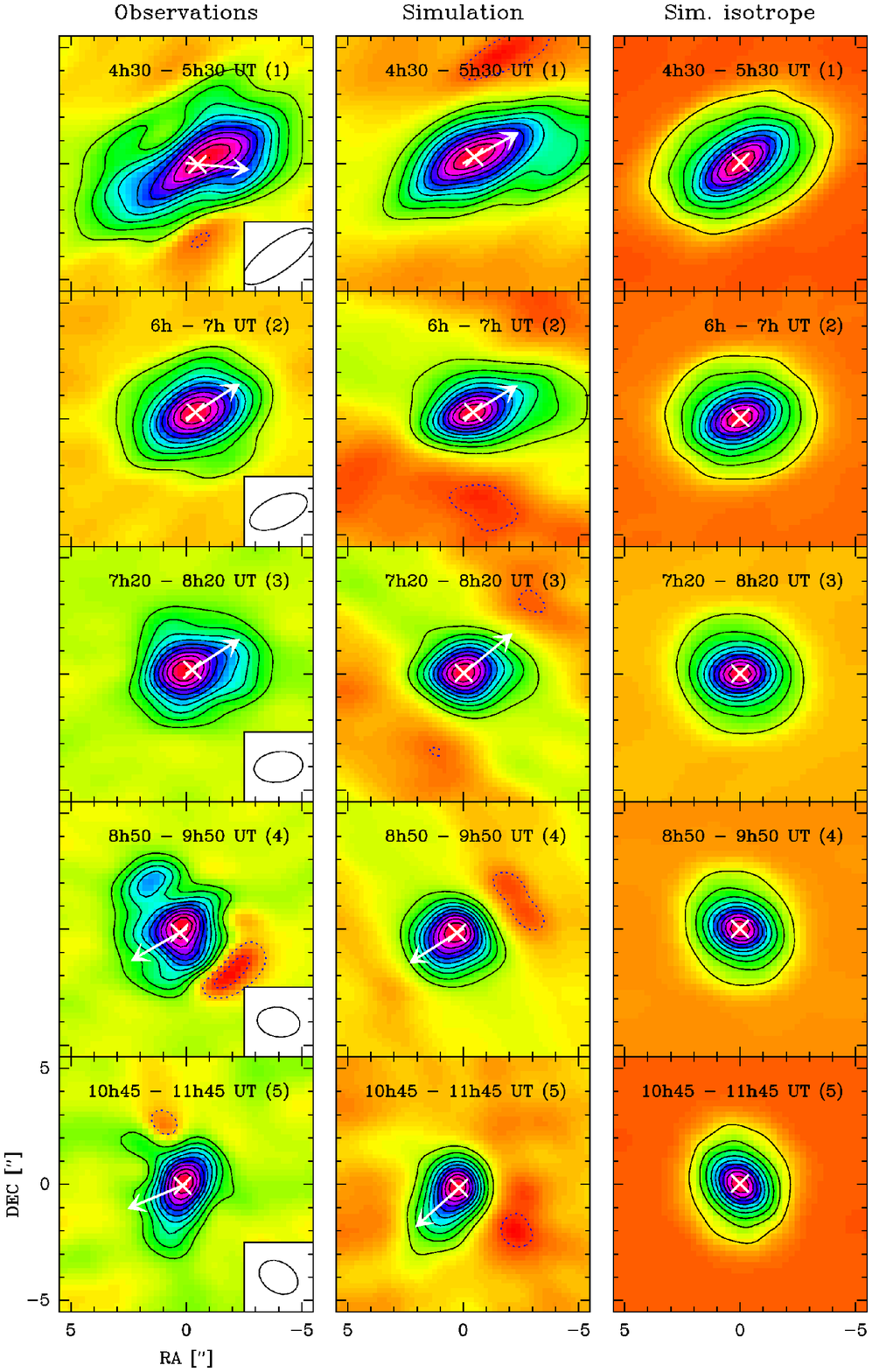}} 
\caption{CO 230 GHz Hale-Bopp maps as a function of time. Left column: observations; central column: simulations with the jet and rotation axis 
parameters which reproduce at best the observations (set (3): \thw~=~80\degre, 
\paw~=~210\degre, $\Psi$~=~18.3\degre, \lat~=~20\degre and \fco~=~35.5\%); 
right column: simulations for an isotropic coma.} 
\label{evol-cartes} 
\end{figure} 
 
\begin{figure} 
\resizebox{\hsize}{!} 
{\includegraphics{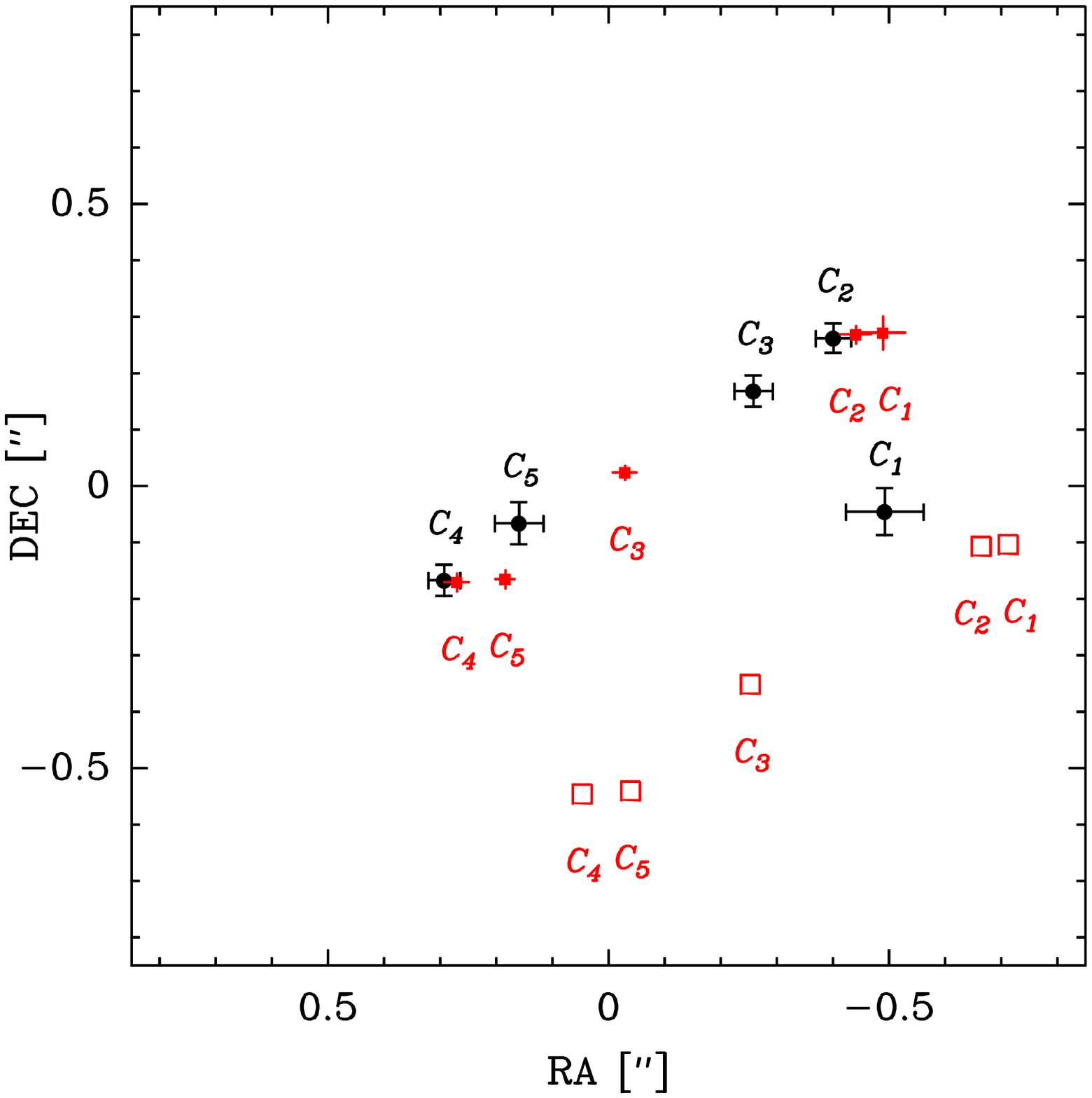}} 
\caption{Time evolution of the photometric centre, as determined 
from fitting in the $uv$-plane. Measurements on Hale-Bopp 
CO 230 GHz observations are shown by black dots with error bars. Those 
from simulations with jet model (3) (\paw~=~210\degre) are plotted  
with red squares: empty squares show absolute positions; filled 
squares show photometric centres with respect to 
the mean photometric center over the period, which can be directly compared 
to the real data. The accuracy of the photometric centre position measured  
on the simulated data is also shown.} 
\label{evol-centre} 
\end{figure} 
 
\subsection{Maps} 
 
Simulated maps as a function of time are compared to observed maps in 
Fig.~\ref{evol-cartes}. The shape of the observed CO coma is relatively 
well reproduced by the model. Some differences may be due to the presence 
of other CO coma features, as suggested by noticeable discrepancies at 
8h50--9h50 UT. The time evolution of 
the photometric centre measured on the simulated data (jet parameters (3) with \paw~=~210\degre) is shown in 
Fig.~\ref{evol-centre}.  When the position of 
the photometric centre is refered to the mean photometric 
centre for the observing period, and is therefore directly 
comparable to the measurements, there is good agreement in the overall 
evolution. As could be expected, the relative 
(modelled$-$observed) positions generally differ, with discrepancies 
reaching $\sim$0.3\arcsec~for 4h30--5h30 and 7h20--8h20 data (maps 1 and 3). 
However, the good overall agreement confirms a posteriori that the 
observed time evolution of the 
CO 230 GHz peak brightness position is related to the CO rotating coma. 
 
Synthetic 230 GHz spectral maps directly comparable to the 
observations (Fig.~\ref{co21-25}) were computed. Some observed 
basic features are well reproduced, as the very peaked and strong 
emission in the blue channels at high velocity and the elongated 
coma in East-West direction for the blue channels at velocities 
close to zero. However, the asymmetry of the spatial distribution 
observed in the red channels is not obtained in the model. As a 
matter of fact, the model, which parameters were constrained from 
the large scale ON--OFF observations, predicts that the inner 
parts of the jet probed during the course of the interferometric 
observations were most of the time projecting Earthward (i.e., 
with negative Doppler velocities). The discrepancies between 
modelled and observed spectral visibilities discussed previously 
are observed directly on the spectral maps. 
 
\subsection{Evidence for a second moving structure} 
\label{otherstructure} 
 
\begin{figure*} 
{\includegraphics[angle=270,width=17cm]{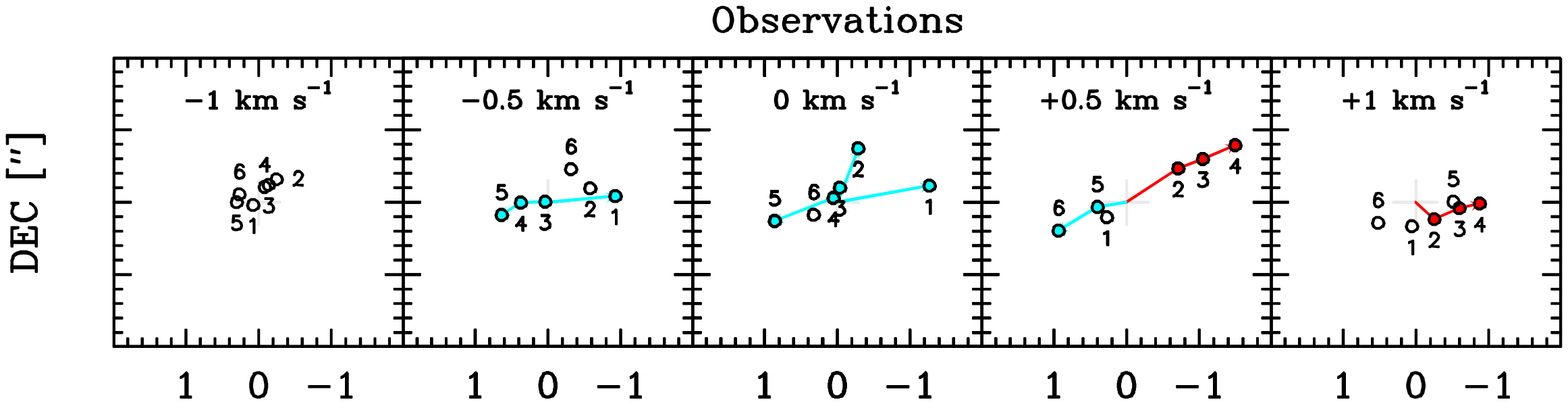}}
{\includegraphics[angle=270,width=17cm]{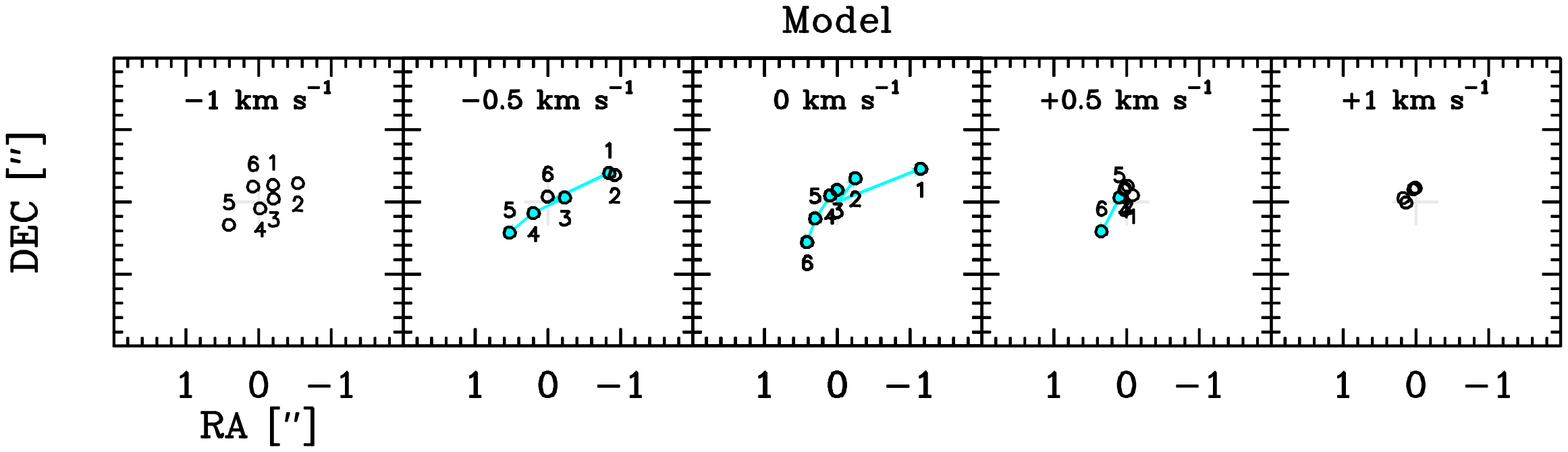}} 
\caption{Time evolution of the CO 230 GHz photometric centre for 
individual velocity channels at $v$ = --1.0, --0.5, 0, +0.5, +1.0 
km s$^{-1}$. The spectral data were smoothed to a spectral 
resolution of 0.5 km s$^{-1}$. The time intervals labelled $1$ to 
$6$ are given in Sect.\ref{otherstructure} . Note that they do not 
correspond to those used for Figs.~\ref{evol-cartes} and 
\ref{evol-centre}. The top figure shows the measurements, while 
the model (set of parameters (3) in Table~\ref{selected-couples}) 
is shown in the bottom figure. In order to illustrate how the 
photometric centres are affected by the main jet, we have 
connected points (coloured in blue) indicative of the motion of 
the CO main jet. The so-called ''red jet'' is shown by red 
symbols.} \label{otherjet} 
\end{figure*} 
 
The inability of our one-jet model to reproduce satisfactorily the 
interferometric data is due to the presence of a second moving CO 
structure, possibly produced during an outburst. 
 
This moving structure is that seen North-West from the nucleus at 
positive velocities in the time averaged channel maps (Fig.~\ref{co21-25}). 
This structure, which is moving away from the observer, is oriented along 
the fringes of baselines 1--4, 1--5, 
3--5 and 4--5, i.e., along those for which a strong discrepancy 
in the time variation of the velocity shift is observed. This 
structure is essentially detected by these baselines : hence, 
the velocity shift measured for these baselines is smaller 
than expected (Fig.~\ref{evol-velo-int-mod}). 
 
In order to study the time evolution of this stucture, the 
spectral data were smoothed to a spectral resolution of 0.5 km 
s$^{-1}$, with velocity channels centred at --1.0, --0.5, 0, +0.5, 
+1.0 km s$^{-1}$ with respect to the comet rest velocity. Five 
consecutive time intervals ($\#1$, $\#$2, $\#$4, $\#$5, $\#$6) of 
$\sim$ 50 min to 1h30 long, were considered. The time intervals 
correspond to characteristic line shapes for the spectrum recorded 
by baseline 3--4 (Fig.~\ref{sp-int}): $\#1$ (4h30--5h48 UT) 
symmetric shape ; $\#$2 ( 5h48--7h07 UT) and $\#$4 (7h07--9h08 UT) 
asymmetric shape with strong blueshift ; $\#$5 (9h08--10h00 UT) 
symmetric shape; $\#$6 (10h00 UT--11h28) asymmetric shape with 
high redshift. A sixth period, labelled $\#$3, merging intervals 
$\#$2 and $\#$4, was also considered. 
 
The 5$\times$6 extracted data sets were analysed in the Fourier 
plane assuming that the brightness distribution is a 2-D Gaussian, 
with its intensity, width and offset with respect to the 
line-integrated, time-averaged centre $C_{m}$ as free parameters. 
Figure~\ref{otherjet} shows the observed time evolution of the 
offset (RA(t), Dec(t)) for each channel. The same plot is shown 
using model (3) as input. 
 
For a non-rotating nucleus, the motion of a cloud of gas released 
during a short time interval from a nuclear source would be seen 
by a straight line in the (RA, Dec) quadrant and the velocity 
channel where its velocity vector projects. For a rotating 
nucleus, a permanent nuclear source near the equator will produce 
an expanding jet that will appear successively in the different 
channels as the nucleus rotates. In the case of comet Hale-Bopp, 
if the equatorial jet projects North-West in (RA, Dec) at positive 
velocities channels ($V_+$) at a given time, then it will 
successively project North-West in negative velocity channels 
($V_-$), South-East/$V_-$, South-East/$V_+$. As shown in 
Fig.~\ref{otherjet}, there is a satisfactory agreement between the 
model and the observations for the channels centred on --1.0, 
--0.5 and 0 km/s. We have tentatively connected the points 
corresponding to the motion of the CO main jet. At time $\#1$, the 
jet is Westward in the $v$ = 0 channel, at time $\#$4 it is moving 
Eastward in the $v$ = --0.5 km s$^{-1}$ channel, at time $\#$5 
Eastward in the $v$ = 0 channel, and at time $\#$6 Eastward in the 
$v$ = +0.5 km s$^{-1}$ channel. 
 
The motion of the red structure is clearly apparent on the 
channels at $v$ = +0.5 and +1 km s$^{-1}$. The direction of the 
motion suggests that it originates from a low latitude region at 
the nucleus surface. The longitude of this source is estimated to 
be within 90--150\degre{} westward from the source of the main 
jet. The weak contribution of the red structure to some baselines 
can be explained by spatial filtering, implying a compact 
structure. The CO 230 GHz flux density in channels $v$ = +0.5 and 
+1 km s$^{-1}$ is four times smaller than in channel $v$ = --0.5 
km s$^{-1}$ at the time the CO main jet is contributing. This may 
explain why the red source does not contribute much to the ON--OFF 
spectra.

\section{Discussion} 
\label{ccl} 
 
\subsection{Model assumptions} 
 
The observations presented in this paper were interpreted with a 
simple geometric model of a CO rotating coma. Besides the assumed 
conical shape of the jet at ejection (see the discussion below), several other 
simplifying assumptions were made to limit the number of free 
model parameters. The outflow velocity was fixed and taken to be 
constant throughout the coma, whereas some acceleration is 
expected by gas dynamics models \citep{combi99}. Day-to-night 
asymmetries in velocity for the CO background gas were not taken 
into account either, though they are clearly present. Indeed, the 
velocity cutoff and width of the blue wing of the CO velocity 
profiles are 10\% higher than the corresponding values for the red 
wings (Fig.~\ref{sp-int}), which suggests higher velocities 
towards the sunlit side of the nucleus given the Earth-comet-Sun 
geometry. In addition, we assumed that CO molecules inside the jet 
expand at the same velocity than those in the background. We 
believe that including slight variations in the flow velocity 
field would not change the main conclusions of this paper, though 
the characteristics of the jet could somewhat differ. For example, 
including this day-to-night asymmetry in velocity would shift the 
jet towards lower latitudes. Higher velocities inside the jet 
would require a proportionally higher jet contribution to the 
total CO production \fco to fit the ON--OFF data. We also did not 
consider possible temperature variations in the coma which have 
direct effects on CO excitation. A kinetic temperature higher in 
the jet than in the background gas would have resulted in a higher 
inferred \fco. However, a 10 K difference in temperature would 
change \fco by less than 10\%.

\subsection{CO jet} 
 
Our observations, interpreted with the help of a simple model, 
suggest the presence of a spiralling "jet" of CO, originating from the 
nucleus. 
The rotation of this jet is consistent with the rotation 
period and axis direction derived from most optical studies. The jet 
originates from a low-latitude ($\approx$ 20\degre) region of the nucleus, and comprises 
a significant amount of the total CO production ($\sim$ 40\%) in a $\sim$20--30\degre wide 
aperture. This is 
the first evidence of a CO spiralling coma around a comet nucleus. 
CO ro-vibrational line emissions observed in comet Hale-Bopp from January to May 1997 show strong 
East/West asymmetries, with the position of the maximum brightness 
moving with time from up to 
$\sim$2\arcsec~West to $\sim$2\arcsec~East of the continuum peak \citep{disanti01,bro03}. This 
is likely caused by the rotating CO structure observed in the radio \citep{bro03} and shows that 
this CO structure was not a transient phenomenon related to some outburst. 
 
Our model, which includes a single jet, appears to be too simple 
to interpret all observation characteristics. Other "jets" are 
present, in particular another moving structure has been 
identified at positive velocities in the interferometric data. Its 
full characterization is difficult from our data and beyond the 
scope of this paper.  The CO main jet is certainly more complex 
than a simple conical spiralling structure. Gas structures created by 
inhomogeneities in gas production at the nucleus surface will  
unescapably be modified during their outflow by the surrounding 
collisional environment \citep[the collisional sphere in Hale-bopp 
coma exceeded 10$^5$ km near perihelion,][]{combi99}. 
 
Gas dynamics models investigating the behavior of expanding dusty gas jets have been developed 
\citep[see the review of][]{Crifo05}. One or several active sources releasing water were 
considered, the rest of 
the nucleus being assumed to be weakly active. 
They show that, due to lateral expansion driven by pressure gradients, jets broaden 
with increasing distance to nucleus and interact with each other, to become eventually almost 
indiscernible in the outer coma. In addition, if there is solar modulation of the surface 
gas flux, the jet morphology will change its appearance as the nucleus rotates.  So far, 
these calculations may not be relevant to the CO jet seen in comet Hale-Bopp. First of all, 
it is very likely that CO diffuses from subsurface layers. Our observations suggest that, in first 
order, there 
is no significant day-to-night asymmetry in the amount of CO gas released by the active source. 
Second, if the active source responsible for the CO jet does not produce similar strong enhancement 
in water production, pressure smoothing will not be acting much and jet broadening may rather result 
from molecular diffusion. In this case, it is possible that a nearly invariant CO density pattern 
co-rotates with the nucleus and remains relatively well preserved at large scales in the coma. 
However, it can be anticipated that the jet development will be still influenced by the 
surrounding environment. 
 
A good illustration is the following. \citet{Rod05} performed 
time-dependent 3-D gas dynamics calculations simulating 
rotationally-induced gas coma structures. They used an arbitrarily 
Halley-like aspherical nucleus, scaled to Hale-Bopp nucleus size 
and homogeneous in composition. CO was assumed to diffuse 
uniformly from below the surface, while H$_2$O sublimates 
according to solar illumination. The computed CO coma is 
asymmetric. The model produces faint H$_2$O and CO spiralling 
structures resulting from weak shocks induced by the surface 
topography. In other words, the CO outflow is largely influenced 
by the general and detailed properties of the flow. The analysis 
of the CO Plateau de Bure observations using time-dependent 
(multi-fluid) gas dynamics calculations was performed by 
\citet{Boi05,Boi+09}. They show that the observed time variations 
cannot be explained by the above mentioned shock structures and is 
due to a strong inhomogeneity in CO production from the nucleus 
surface or sub-surface. It is possible to hypothesize lengthly 
about the meaning of this CO overproduction. It can traduce 
inhomogeneities in CO content inside the nucleus but also local 
variations in mantle thickness, or in some properties of the 
nucleus material (e.g., dust/ice matrix structure, thermal 
conductivity). We leave this discussion to experts and encourage 
them to perform numerical simulations \citep[see the review 
of][]{Pria05}.

\subsection{Comparison with other studies} 
 
Two questions arise: 
 
1) Is the CO jet we observed related (correlated) to jets observed for 
other molecules? 
 
2) Is the CO jet we observed related (correlated) to observed dust 
jets? 
 
In their interferometric observations of several molecular species 
(HCN, HNC, DCN, HDO) with the OVRO array, \citet{bla99} observed 
that the molecular emissions peak at positions offset by a few 
arcsec from the continuum emission of the comet, and attributed 
these offsets to molecular jets. We do not observe such large 
offsets for CO when averaging the data over 70\% of the rotation 
period. However, offsets reaching 1--1.5\arcsec{} were observed 
for other molecules observed at the PdBI, interpreted to be the 
gaseous signatures of the high latitude dust jet observed in the 
visible \citep{boi+07}. The strong CO jet has no strong H$_2$S, CS, 
SO and HCN counterparts though weak rotational modulations in the 
line shapes of their radio emission are observed for some of them 
\citep{boi+07}.

\citet{led02} and \citet{led09} observed spiralling jets of OH, CN and C$_2$. 
Several (up to five) active areas were necessary to reproduce 
their observations from their Monte Carlo simulation. The 
strongest jets are coming from low-latitude regions (--22\degr{} 
to +20\degr). One of them could thus be associated with our CO 
jet. One low-latitude southern area is responsible for one-half of 
the OH coming from jet-source, but has a large opening angle 
(120\degr). So it is likely not related to the CO jet. Spatial 
profiles of H$_2$O obtained from long-slit observations of 
infrared water lines only show small East-West asymmetries 
\citep{dello00}. In summary, there is no strong evidence for a 
strong H$_2$O jet associated to the CO jet. 
 
\citet{woo02} mapped HCN in comet Hale-Bopp with the BIMA array. 
Their observations, which spanned over several days, were binned 
and averaged according to the phase of the comet rotation, in 
order to avoid smearing due to rotation. They found a jet 
morphology, possibly correlated with CN imaged in the visible. 
However, there is apparently no correlation of HCN with dust jets. 
The presence of a high-latitude HCN jet is however suggested from 
the PdBI data \citep{boi+07}. 
 
\begin{figure} 
\resizebox{\hsize}{!} 
{\includegraphics{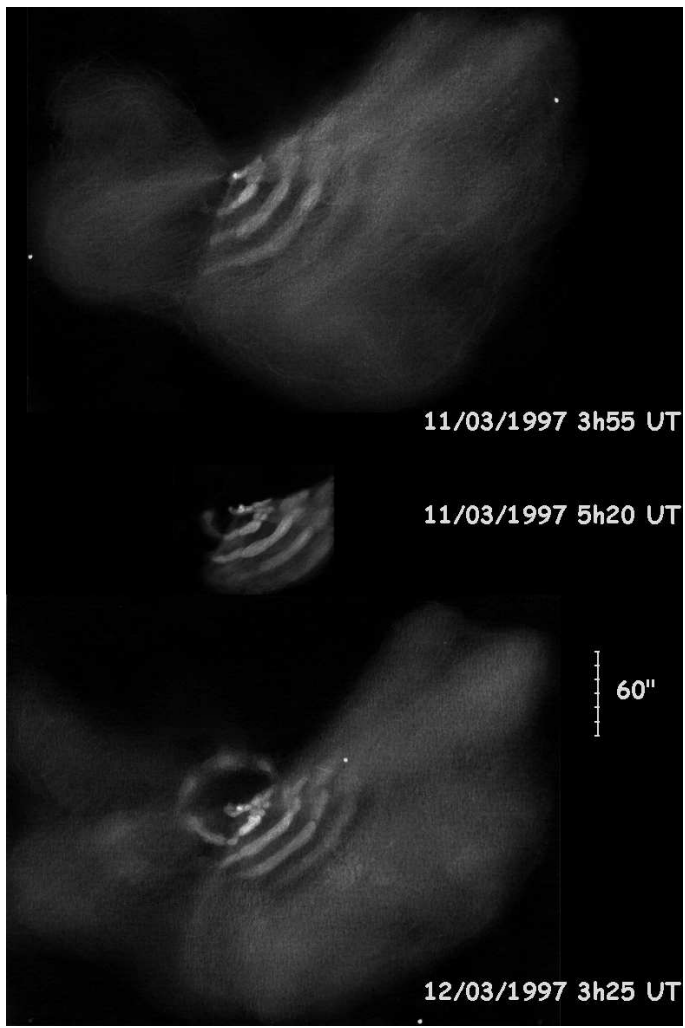}} 
\caption{Drawings of the inner coma of comet Hale-Bopp on 11 and 
12 March, 1997 (N. Biver, 25.6 cm Newtonian telescope).} 
\label{outburst-drawing} 
\end{figure}

Dust jets also appear to be uncorrelated with our CO jet.  Indeed, 
from visible imaging at the time of the PdBI observations, 
\citet{jorda99} observed a high latitude (64\degr) dust jet.  In 
the analysis of \citet{vas99a}, strong dust jets are present from 
latitudes +65\degr and --65\degr, in addition to weaker jets from 
low latitudes \citep[see also][]{sch04}. In February--May 1997, 
the high-latitude northern jet produced repetitive sunward shells 
instead of a full spiral (Fig.~\ref{outburst-drawing}), indicating 
that the source of this dust jet shut off during night time in 
contrast to the CO jet. 
 
In addition to the CO main jet, we have identified a second moving 
structure in red channels that we believe originates from a low 
latitude source. Interestingly, on 12.14 March 1997 UT the dust 
coma presented a well defined shell at $\sim$ 40\arcsec{} in a 
direction opposite to the repetitive sunward shells produced by 
the high-latitude sources (Fig.~\ref{outburst-drawing}). This 
shell was not present on 11.16 March, nor on 14.80 March, and can 
be attributed to an outburst initiated on 11.20 March UT, i.e., at 
the time when the red structure began its expansion. Possibly, the 
red structure is related to this outburst. 
 
\subsection{Plans for further studies} 
 
ON--OFF and interferometric data were obtained at the Plateau de 
Bure interferometer for HCN, HNC, CS, H$_2$S, SO, H$_2$CO and 
CH$_3$OH \citep{wink99,boi+07}. Some lines (e.g., lines of HCN and 
CS) do show rotation-induced variations in their velocity shifts. 
The analysis of the Plateau de Bure observations is continuing and 
will be presented in forthcoming papers. 
 
The present study showed that radio observations can provide 
valuable information on the distribution of parent molecules in 
inner cometary atmospheres and its temporal evolution. In contrast 
with standard imaging techniques, radio observations are sensitive 
to radial velocities, i.e. they are sensitive to the gas 
distribution along the line of sight, whereas the other techniques 
are rather sensitive to the distribution on the plane of the sky. 
They also probe different gas species. Radio observations and 
other techniques are therefore complementary. 
 
Radio interferometric imaging is a powerful tool for astrometry. 
Our observations show that, in addition, rotating comas can be 
detected from the motion of the centroid of molecular maps. This 
opens new perspectives because possibly useful constraints on the 
rotation properties of cometary nuclei will be obtained from such 
measurements. 
 
The analysis of our interferometric data was hampered by to the 
limited instantaneous $uv$-coverage of the Plateau de Bure 
interferometer. The Atacama Large Millimeter and submillimeter 
Array (ALMA), with its 50 antennas, will be able to obtain images 
of molecular and continuum emissions with a short sampling time, 
high sensitivity and high angular resolution. It will provide a 
3-D dynamical picture of inner cometary gaseous atmospheres, 
simultaneous images of the dust coma, and spatial information on 
the gas temperature \citep{biver05,bocalma08}. Important 
breakthroughs concerning nucleus and coma processes can be 
expected.

\begin{acknowledgements} 
This paper is dedicated to J\"{o}rn Wink, who performed these 
marvellous observations and helped us in their analysis despite 
his terrible illness. We gratefully thank the IRAM staff, for 
help in the observations, and P. Rocher 
(IMCCE, Observatoire de Paris) for providing us with detailed 
ephemeris of comet Hale-Bopp. We thank Jean-Fran\c{c}ois 
Crifo for enlightening discussions about coma hydrodynamics. Many thanks 
also to for Anne Dutrey for 
constant support, and to Laurent Jorda, for 
helpful exchanges about comet Hale-Bopp rotation. 
IRAM is an international institute co-funded by the CNRS, France, the 
Max-Planck-Gesellschaft, Germany, and the Instituto Geogr\`afico Nacional, 
Spain. This work has been supported 
by the Programme national de plan\'etologie of Institut national des 
sciences de l'univers. 
 
\end{acknowledgements} 
 
\bibliographystyle{apj}

\end{document}